\documentclass[iop,twocolumn]{aastex631}
\usepackage{amsmath}

\usepackage{gensymb}
\usepackage{hyperref}
\usepackage{graphicx,latexsym}
\newcommand{\abs}[1]{\left\lvert#1\right\rvert}
\newcommand\xone{x_1}
\newcommand\xtwo{x_2}
\newcommand\xthree{x_3}

\newcommand\pcc{\;{\rm cm}^{-3}}
\newcommand\Msun{\; {\rm M}_{\odot}}

\newcommand\kms{\; {\rm km}\;{\rm s}^{-1}}
\newcommand\pc{\;{\rm pc}}

\newcommand\kpc{\;{\rm kpc}}
\newcommand\Mpc{\;{\rm Mpc}}

\newcommand\erg{\; {\rm erg}}

\newcommand\yr{\; {\rm yr}}
\newcommand\Myr{\;{\rm Myr}}
\newcommand\Gyr{\;{\rm Gyr}}
\newcommand\sfrunit{{\Msun \yr^{-1}}}

\newcommand\Surf{\Msun\;{\rm pc^{-2}}}

\newcommand\dunits{\Msun\;{\rm pc^{-3}}}

\newcommand\C{{\mathcal C}}

\newcommand\simgt{\lower.5ex\hbox{$\; \buildrel > \over \sim \;$}}
\newcommand\simlt{\lower.5ex\hbox{$\; \buildrel < \over \sim \;$}}
\newcommand{\RNum}[1]{\uppercase\expandafter{\romannumeral #1\relax}}

\def\spose#1{\hbox to 0pt{#1\hss}}
\def\dt{\spose{\raise 1.0ex\hbox{\hskip2pt$\mathchar"201$}}}

\shortauthors{Li et al.}

\begin{document}

%\title{How Gas Flows Under Two Bars} 
\title{How Nested Bars Enhance, Modulate, and are Destroyed by Gas Inflows}

\author[0000-0002-0627-8009]{Zhi Li}
%\email{lizhi@shnu.edu.cn}
\affiliation{Shanghai Key Lab for Astrophysics, Shanghai Normal University, 100 Guilin Road, Shanghai 200234, P.R. China}
\affiliation{Department of Astronomy, School of Physics and Astronomy, Shanghai Jiao Tong University, 800 Dongchuan Road, Shanghai 200240, P.R. China}
\affiliation{Tsung-Dao Lee Institute, Shanghai Jiao Tong University, Shanghai 200240, P.R. China}

\author[0000-0001-9953-0359]{Min Du}
\correspondingauthor{Min Du}
\email{dumin@xmu.edu.cn}
\affiliation{Department of Astronomy, Xiamen University, Xiamen, Fujian 361005, P.R. China}

\author[0000-0001-7902-0116]{Victor P. Debattista}
%\email{vpdebattista@uclan.ac.uk}
\affiliation{Jeremiah Horrocks Institute, University of Central Lancashire, Preston PR1 2HE, UK}

\author[0000-0001-5604-1643]{Juntai Shen}
\correspondingauthor{Juntai Shen}
\email{jtshen@sjtu.edu.cn}
\affiliation{Department of Astronomy, School of Physics and Astronomy, Shanghai Jiao Tong University, 800 Dongchuan Road, Shanghai 200240, P.R. China}
\affiliation{Key Laboratory for Particle Astrophysics and Cosmology (MOE) / Shanghai Key Laboratory for Particle Physics and Cosmology, Shanghai 200240, P.R. China}

\author[0000-0002-1253-2763]{Hui Li}
%\email{}
\affiliation{Department of Astronomy, Tsinghua University, Beijing 100084, P.R. China}
\affiliation{Department of Astronomy, Columbia University, New York, NY 10027, USA}
%\affiliation{Department of Astronomy, Columbia University, New York, NY 10027, USA}
%\affiliation{Department of Physics and Kavli Institute for Astrophysics and Space Research, Massachusetts Institute of Technology, Cambridge, MA 02139, USA}

\author[0000-0003-4052-8785]{Jie Liu}
\affiliation{Shanghai Astronomical Observatory, Chinese Academy of Science, 80 Nandan Road, Shanghai, 200030, P.R. China}

\author[0000-0001-8593-7692]{Mark Vogelsberger}
\affiliation{Department of Physics \& Kavli Institute for Astrophysics and Space Research, Massachusetts Institute of Technology, Cambridge, MA 02139, USA}

\author[0000-0002-8658-1453]{Angus Beane}
\affiliation{Center for Astrophysics $|$ Harvard \& Smithsonian,  Cambridge, MA, USA}

\author[0000-0003-3816-7028]{Federico Marinacci}
\affiliation{Department of Physics \& Astronomy ``Augusto Righi", University of Bologna, via Gobetti 93/2, 40129 Bologna, Italy}

\author[0000-0002-3790-720X]{Laura V. Sales}
\affiliation{Department of Physics and Astronomy, University of California, Riverside, CA, 92521, USA}

%\author[0000-0003-4052-8785]{all other SMUGGLE team members}
%\affiliation{XXX}

%===============================================================================

\begin{abstract}

Gas flows in the presence of two independently-rotating nested bars remain not fully understood, which is likely to play an important role in fueling the central black hole. We use high-resolution hydrodynamical simulations with detailed models of subgrid physics to study this problem. Our results show that the inner bar in double-barred galaxies can help drive gas flow from the nuclear ring to the center. In contrast, gas inflow usually stalls at the nuclear ring in single-barred galaxies. The inner bar causes a quasi-periodic inflow with a frequency determined by the difference between the two bar pattern speeds. We find that the star formation rate is higher in the model with two bars than in that with one bar. The inner bar in our model gradually weakens and dissolves due to gas inflow over a few billion years. Star formation produces metal-rich/$\alpha$-poor stars which slows the weakening of the inner bar, but does not halt its eventual decay. We also present a qualitative comparison of the gas morphology and kinematics in our simulations with those of observed double-barred galaxies.

\end{abstract}


%-----------
%-- Sect. 1
%-----------

\section{Introduction}
\label{sec:intro}

Galaxies with two nested stellar bars are termed double-barred galaxies \citep[S2Bs, see][]{laine_etal_02,erwin_04,shlosm_05,buta_etal_15}. About 20\% of disk galaxies appear to be S2Bs \citep{laine_etal_02,erwin_11}. In such systems the two bars are observed to be randomly oriented with respect to each other \citep[][]{but_cro_93,fri_mar_93}, implying two independent rotation speeds that are consistent with theoretical expectations \citep[e.g. ][]{mac_spa_00,deb_she_07,du_etal_15}. Inner bars in S2Bs are also dubbed as ``nuclear'' or ``secondary'' bars in some works. These terms may be less appropriate since observations have shown that inner bars can be as large as some short single bars \citep[semi-major axes of $2$-$3\kpc$, see][]{erwin_05,lorenz_etal_20}. Recent studies have suggested that inner bars are probably scaled-down replicas of outer bars based on resolved spatial maps of the mean stellar population properties \citep[][]{mendez_etal_19,lorenz_etal_19b,bittner_etal_21}.

The gas inflow driven by the large-scale bar generally stalls at a starburst nuclear ring with a typical size of a few hundred parsecs, preventing accretion onto the supermassive black hole \citep[SMBH, e.g. ][]{fanali_etal_15,li_etal_17,tress_etal_20}. Inner bars have been hypothesized to be an important mechanism for further removing gas angular momentum, in a manner similar to their large-scale counterparts \citep[e.g. ][]{shlosm_etal_89, fri_mar_93}. The inflowing gas driven by the inner bar may be a possible mechanism to contribute the growth of the SMBH in the galactic center. However, the fuel may be choked off as the inner bar gradually dissolves for a short time ($\sim1\Gyr$) when the SMBH becomes massive enough \citep[$\sim0.1\%$ stellar mass of the host galaxy, ][]{du_etal_17,nak_bab_23}, and the growth is therefore self-limiting. The remnant of the destroyed inner bar leads to the formation of a spheroidal component that is similar to a small classical bulge \citep[][]{guo_etal_20}. Interestingly, observations have found local S2Bs are likely to host classical instead of pseudo bulges \citep[][]{lorenz_etal_19a}, although it would be difficult to determine the origin of these classical bulges.

One important assumption in the above scenario is that \textit{inner bars promote gas inflows from the nuclear ring to the center}. This has long been speculated to be the case \citep[e.g. ][]{shlosm_etal_89, heller_etal_06, nameka_etal_09, hop_qua_10}, but studies have provided conflicting results \citep[e.g. ][]{maciej_etal_02, rautia_etal_02}. For example, the `bars-within-bars' model of \citet{shlosm_etal_89} and \citet{hop_qua_10} supports the idea that inner bars (or small-scale bars) can reduce gas angular momentum just like large-scale bars; however, some theoretical studies of gas motions in S2B potentials have argued that the gas flow patterns differ fundamentally between single bars and nested bars, and the inner bar cannot enhance the central mass inflow rate due to the lack of shocks \citep[][]{maciej_etal_02,rautia_etal_02,shl_hel_02}. 

Nevertheless, there are still large uncertainties/simplification in the modelling of nested bars that can affect the gas behaviours. For instance the two bars in the S2B simulations of \citet{maciej_etal_02} and \citet{nameka_etal_09} are modeled as rigid (Ferrers) ellipses whose torque distributions are different from those of real bars \citep[][]{but_blo_01,li_etal_17}. In addition, both the pattern speeds and the shapes of the bars are pulsating in a self-consistent, dynamically decoupled double-barred system \citep[e.g. ][]{she_deb_09,du_etal_15,wu_etal_16}, and this effect on gas flows has been less explored in the literature. Moreover, star formation and feedback in the central kiloparsec may also affect the inner bar dynamics, including the gas inflow \citep[][]{woznia_15}, and this requires detailed modeling of sub-grid physics as well as sufficient numerical resolution. It remains an important question, therefore, whether inner bars are able to drive gas inflows and potentially activate the central SMBH, which motivates us to perform an in-depth numerical study of gas flows in S2Bs with state-of-the-art subgrid physics implemented. We would like to investigate: (1) whether the inner bar can drive gas further to the center using a realistic S2B model that qualitatively matches the observed stellar kinematics; (2) how the gas flows respond to a pulsating inner bar; and (3) how different the gas flow pattern is between single- and double-barred galaxies, especially in the central region.

The paper is organized as follows: Section~\ref{sec:sims} describes our simulation setups. Section~\ref{sec:result} presents the evolution of our S2B models using different gas prescriptions and isolates the effects of the inner bar with control models. Section~\ref{sec:compobs} compares the properties of our simulated S2Bs with observations. We discuss the implications and limits of this study in Section~\ref{sec:discussion}. Section~\ref{sec:summary} gives a summary of our results.

%-----------
%-- Sect. 2
%-----------

\section{Simulations}
\label{sec:sims}

The fiducial S2B model of \citet{du_etal_17} nicely reproduces many observed features in real S2B galaxies, such as $\sigma$-humps/hollows and $h_4$ rings\citep[see ][]{du_etal_16,lorenz_etal_19b}. The long-lived inner bar in this model forms spontaneously from a cool inner disk via dynamical instabilities. For our study we choose to adopt a similar setup, as detailed below. We select the snapshot corresponding to the time $t=2.7\Gyr$ of the fiducial model in \citet{du_etal_17} as the initial condition for the stars in our simulations. At this time, the semi-major axis of the inner bar is $\sim0.75\kpc$, while the outer bar has a semi-major axis of $\sim7.5\kpc$. The inner bar rotates roughly three times as fast as its outer counterpart. The two bars have reached a relatively steady state, but the pattern speeds of the two bars are not stationary and the shape of the bars also changes, i.e. the bars are pulsating during the evolution \citep[see also ][]{deb_she_07}. The stellar disk in our model has a mass of $6.0\times10^{10}\Msun$. We use the same logarithmic rigid dark matter halo that \citet{du_etal_15,du_etal_17} used, which has a potential given by $\Phi(r)=0.5V_h^2\ln{(r^2+r_h^2)}$ with $V_h=157.4\kms$ and $r_h=37.5\kpc$, where $r$ is the spherical radius with respect to the galactic center. This rigid halo simplifies our simulation, while the dynamics at the center where the baryonic components dominate, are barely changed. We refer the reader to \citet{du_etal_15,du_etal_17} for further details of the model. 

%fig1
\begin{figure}[!t]
\includegraphics[width=0.4\textwidth]{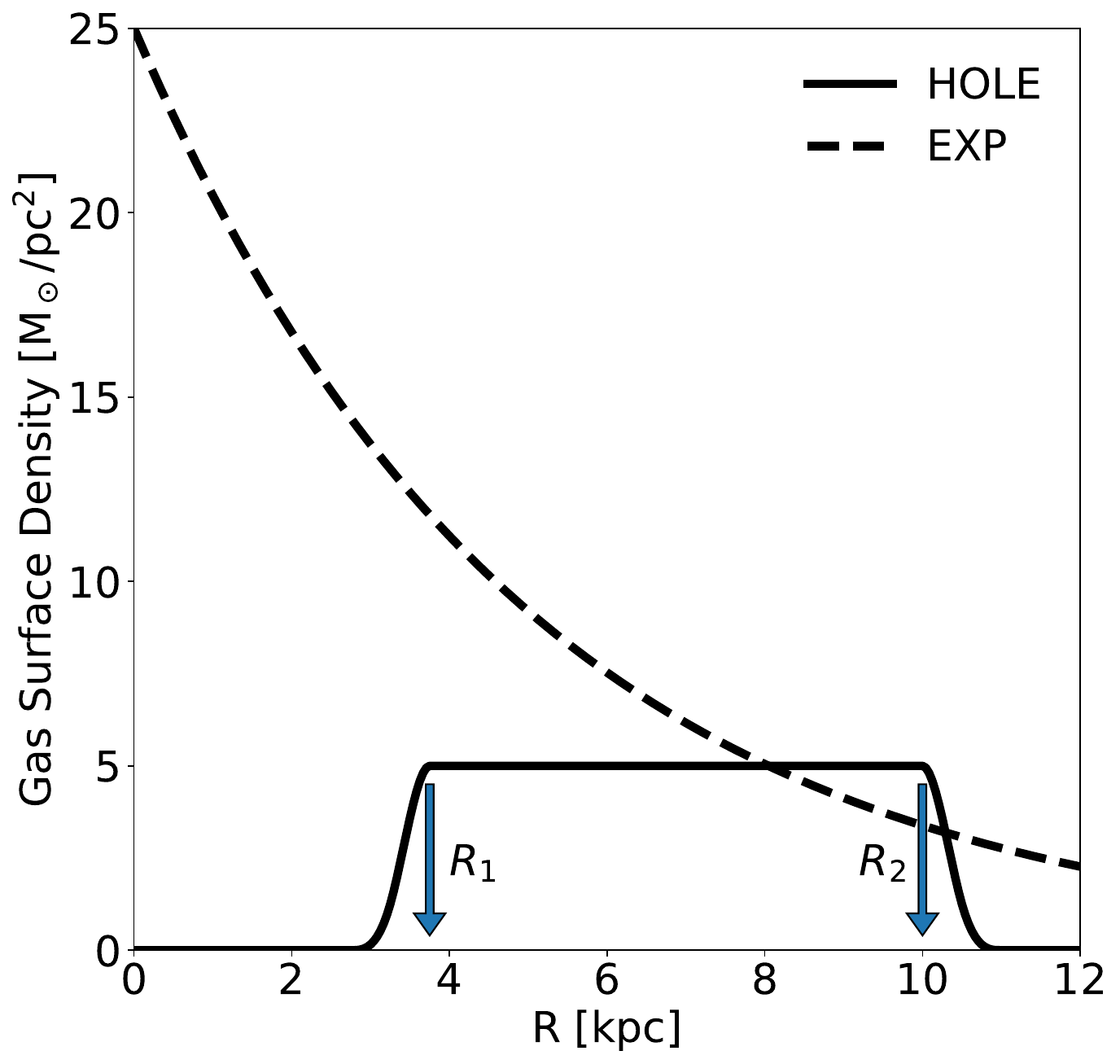}
\caption{Initial gas surface density profiles used in our models. The solid curve shows the `HOLE' profile, with the two arrows denoting the inner and outer boundary at $R_1=3.75\kpc$ and $R_2=10.0\kpc$; the dashed line shows the `EXP' profile with the central density $\Sigma_0=25\Surf$ and the scale radius $R_d=5\kpc$.
\label{fig:initgas}}
\vspace{0.2cm}
\end{figure}

We insert a lightweight gas disk into this snapshot to study gas flows under the live S2B potential. We start the simulation time from when we insert the gas disk. There are two different gas surface density profile at $t=0$ used in our models as shown in Figure~\ref{fig:initgas}.: one is set to an annulus with inner and outer radii at $R_1=3.75\kpc$ and $R_2=10.0\kpc$, respectively, where $R$ is the cylindrical radius with respect to the galactic center (labeled as `HOLE' profile). $R_1$ is chosen to be close to the middle of the outer bar. This annulus configuration aims to protect the inner bar from the rapid formation of a gaseous central massive concentration (CMC) that may dissolve the inner bar quickly \citep[e.g.][]{woznia_15,du_etal_17,guo_etal_20,nak_bab_23}. The initial inner boundary $R_1$ also help us focusing on the inflowed gas during the evolution of the model. The surface density of the annulus gas disk between $R_1$ and $R_2$ is uniform\footnote{A uniform gas disk is commonly used in studying bar-driven gas flows because it results in no pressure gradient in the radial direction \citep[e.g. ][]{athana_92b,eng_ger_97,kim_etal_12a,sorman_etal_15a}.} with $\Sigma_{\rm gas}=5\Msun/{\rm pc}^2$ and is tapered off by a cubic spline function with a characteristic width of $0.2\kpc$ outside $R_1$ and $R_2$ \citep[see also a similar initial gas disk setup in][]{beane_etal_23}. The total gas mass is therefore $1.5\times10^9\Msun$; the other is set to have an exponential surface density profile (labeled as `EXP' profile). The scale radius of the exponential gas disk is $R_d=5\kpc$ and the central gas surface density is $\Sigma_0=25\Surf$. The total gas mass is therefore $3.9\times10^9\Msun$. Although this gas disk changes both the gas surface density profile and the gas mass at the same time, it primarily tests the effects of a massive CMC on bars in a relatively extreme case. The vertical density profile is chosen to be in hydrostatic equilibrium following \citet{spr_her_05} and \citet{wang_etal_10} for both gas disk types. We start the gas disk on circular orbits obtained from the axisymmetrized galactic potential. The initial temperature of the gas disk is set to $1.5\times10^4{\;{\rm K}}$. All gas cells at $t=0$ have solar metallicity and alpha-elements as described in \citet[][]{asplun_etal_09}.

\begin{deluxetable*}{ccccccc}
\tablecaption{Model Setup}
\tablenum{1}

\tablehead{\colhead{Model}     & 
\colhead{$\gamma$}             & 
\colhead{Gas Self-Gravity}     & 
\colhead{\tt SMUGGLE}          & 
\colhead{Gas Profile}          &
\colhead{Gas Fraction}          &
\colhead{Galaxy Type}          \\
\colhead{} & \colhead{} & \colhead{} & \colhead{} 
& \colhead{} & \colhead{} & \colhead{}} 
%% All data must appear between the \startdata and \enddata commands
\startdata
S2BISO & $1.0$ & Off & Off & HOLE & $2.5\%$ & S2B \\
S2BSMU & $5/3$ & On  & On  & HOLE & $2.5\%$ & S2B \\
S1BSMU & $5/3$ & On  & On  & HOLE & $2.5\%$ & SB \\
S0BSMU & $5/3$ & On  & On  & HOLE & $2.5\%$ & SA \\
S2BEXP & $5/3$ & On  & On  & EXP  & $6.5\%$ & S2B \\
\enddata

\tablecomments{List of models in this work. Column 1: model names; Column 2: adiabatic index $\gamma$, $\gamma=1$ corresponds to the isothermal EoS while $\gamma=5/3$ is for ideal monoatomic gas; Column 3: switch for gas self-gravity; Column 4: switch for {\tt SMUGGLE} model; Column 5: initial surface density profile of the gas disk, HOLE means an initial uniform gas disk with an inner boundary of $R_1=3.75\kpc$ and outer boundary of $R_2=10.0\kpc$, while EXP means an initial exponential gas disk without the inner boundary (see also Figure~\ref{fig:initgas}); Column 6: gas mass fraction relative to the stellar disk. Column 7: galaxy morphology of the model. }
\label{tbl:models}
\end{deluxetable*}

Gas cooling and heating, star formation, and stellar feedback are included via the ISM and stellar feedback model model {\tt SMUGGLE} \citep[Stars and MUltiphase Gas in GaLaxieEs model,][]{marina_etal_19}. This model reproduces well the observed feedback-regulated star formation and has been widely used for studying galaxy formation and evolution \citep[e.g. ][]{burger_etal_22,lih_etal_22,sivasa_etal_22,tacche_etal_22}. We adopt a local star formation efficiency (SFE) $\epsilon_{\rm sf}=0.01$ and a star formation density threshold $\rho_{\rm th}=100\pcc$ in {\tt SMUGGLE}. We use the same stellar yields and evolution model as the original Illustris simulation \citep[][]{vogels_etal_13}.

%fig2
\begin{figure*}[!t]
\includegraphics[width=1.00\textwidth]{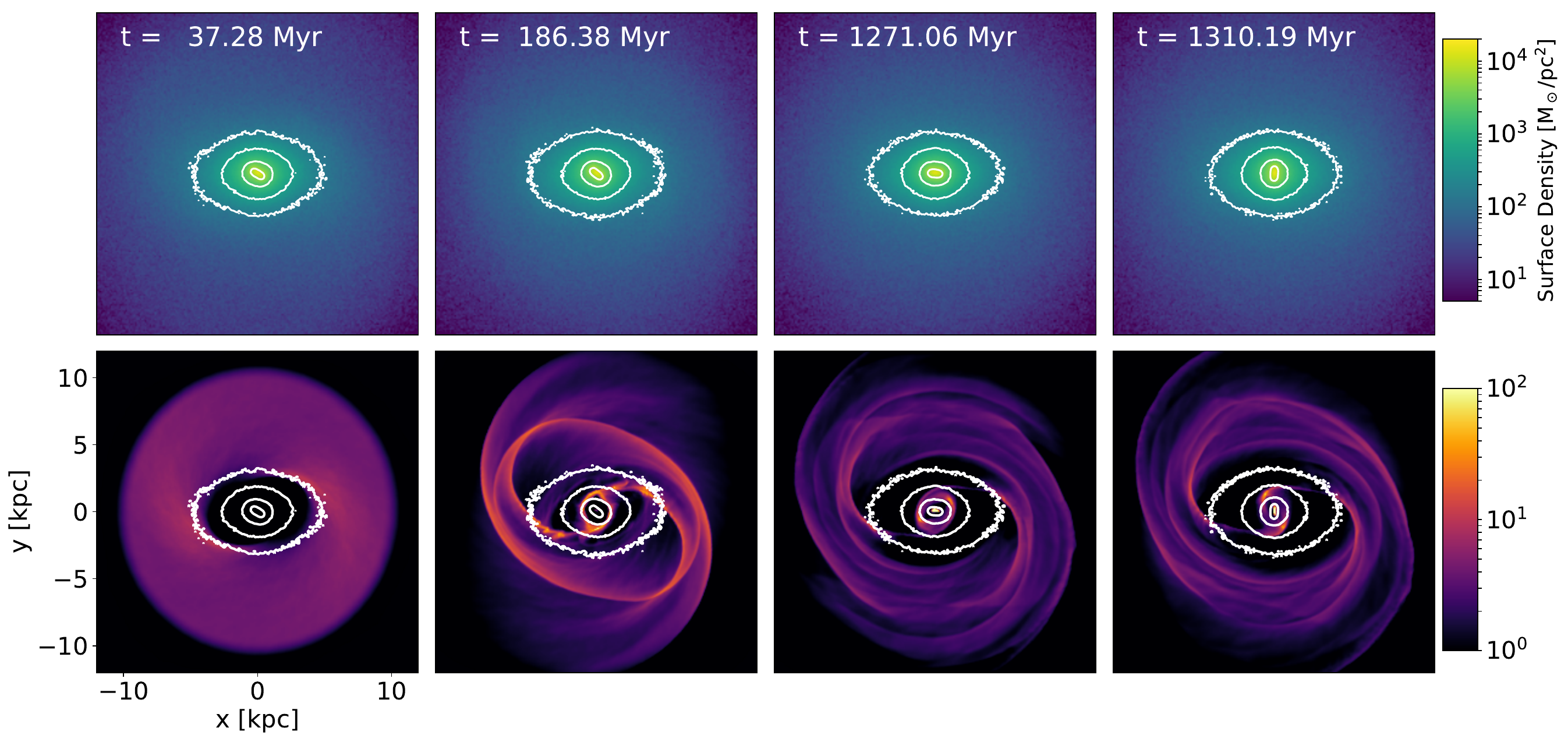}
\caption{Evolution of model S2BISO. Top: stellar surface density. Bottom: gas surface density. White lines represent the contours of the stellar surface density placed at $0.16, 0.54, 1.85$, and $6.31\times10^3\Surf$. The two stellar bars are clearly outlined by the contours. The first two columns show the early evolution when the two bars are misaligned, while they are parallel in the third column and perpendicular in the fourth column. The evolution time is labeled at the top. The disk rotates counterclockwise. 
\label{fig:S2Bisoden}}
\vspace{0.2cm}
\end{figure*}

%fig3
\begin{figure*}[!t]
\includegraphics[width=1.00\textwidth]{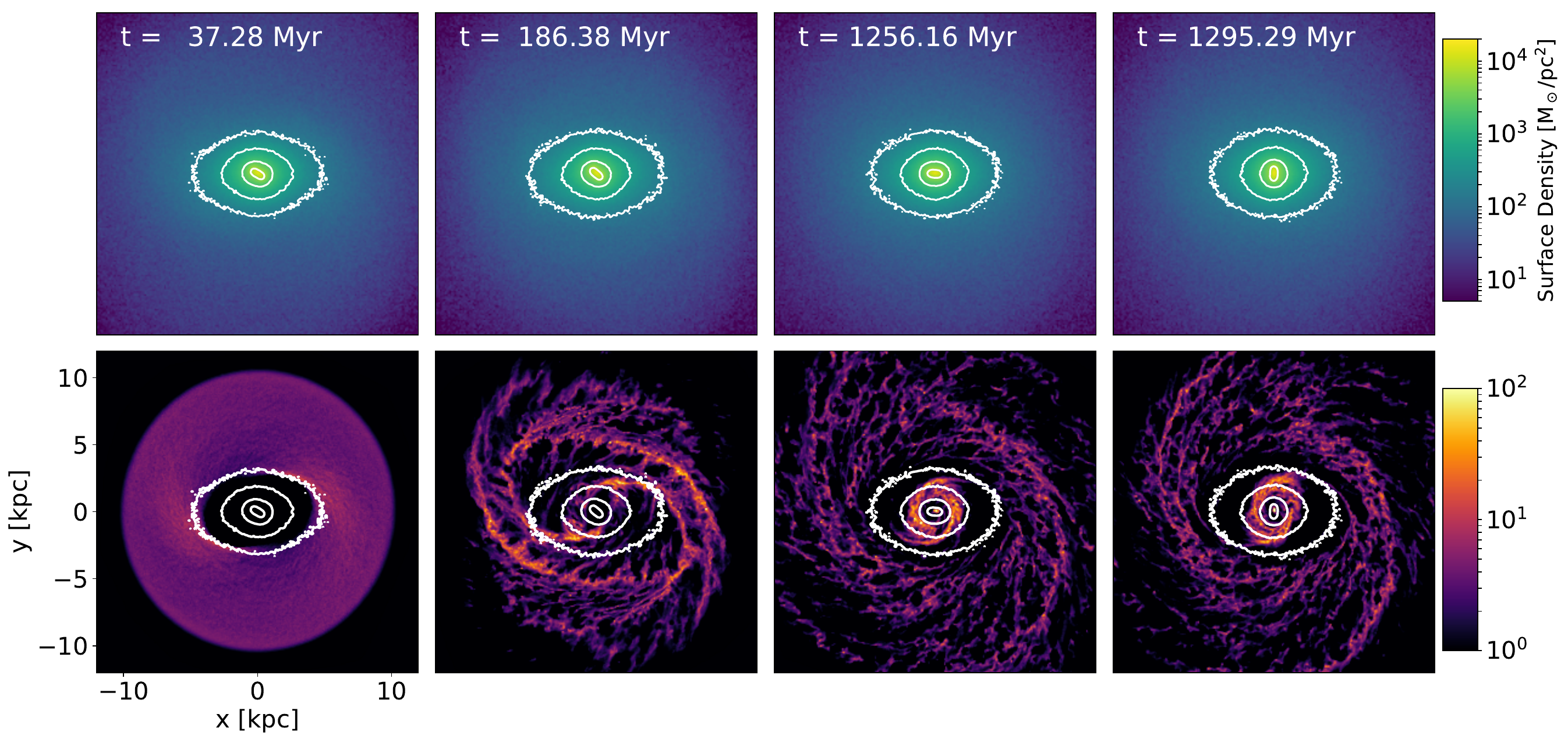}
\caption{Similar to Figure~\ref{fig:S2Bisoden} but for S2BSMU. 
\label{fig:S2Bsmuden}}
\vspace{0.2cm}
\end{figure*}

The simulations are performed using the moving-mesh code {\tt AREPO} \citep[][]{spring_10,weinbe_etal_20}. The simulation box size is $60\kpc$ so the boundary effects on the gas disk are negligible. There are 3.8 million stellar particles with a fixed softening length of $25\pc$ and 1.3 million initial gas cells with adaptive softening enabled. The minimum adaptive softening is set to $0.25\pc$ and the target mass of gas cells is $\approx1100\Msun$. The galaxy properties modeled with {\tt SMUGGLE} nicely converge at this resolution \citep{marina_etal_19}. The minimum gas cell radius in our simulations reaches $\sim0.14\pc$.

We present 5 models with configurations summarized in Table~\ref{tbl:models}. The fiducial S2B model is referred to as S2BSMU, with {\tt SMUGGLE} enabled. S2BISO has the same initial conditions as S2BSMU but uses the isothermal equation of state (EoS) and excludes gas self-gravity to avoid gravothermal catastrophe. The \textit{effective} sound speed in S2BISO is $c_s\approx10\kms$ (obtained from initial gas temperature). This sound speed reflects the average velocity dispersion between molecular clouds rather than the microscopic temperature of the diffuse gas. We further run two additional control models (S1BSMU and S0BSMU) to highlight the gravitational effects of bars: We generate S1BSMU by azimuthally scrambling\footnote{Scrambling means the in-plane position $(x,y)$ and velocity $(v_x,v_y)$ of each stellar particle is rotated by a random angle $\phi$ with respect to the galactic center. $\phi$ follows a uniform distribution on the interval $[0, 2\pi)$ \citep[see also][]{brown_etal_13,li_etal_14}.} the stellar particles at $R\leq0.75\kpc$ of S2BSMU in the initial condition. S1BSMU is thus a single-barred (SB) galaxy whose bar structure is almost the same as the outer bar of S2BSMU. Similarly, we generate S0BSMU by scrambling stellar particles at $R\leq7.5\kpc$, thus producing an unbarred galaxy. These 4 models have the exactly same azimuthally-averaged mass profile and rotation curve at $t=0$. The gas mass fraction relative to the stellar disk in these 4 models is $2.5\%$; these 4 models are run for $\sim1.9\Gyr$. S2BEXP has nearly identical initial conditions as S2BSMU, with the exception of an initial exponential gas disk rather than a uniform gas disk. Consequently, the gas mass fraction is increased to $6.5\%$; this model is run for $\sim3.8\Gyr$.

%fig4
\begin{figure*}[!t]
\includegraphics[width=1.00\textwidth]{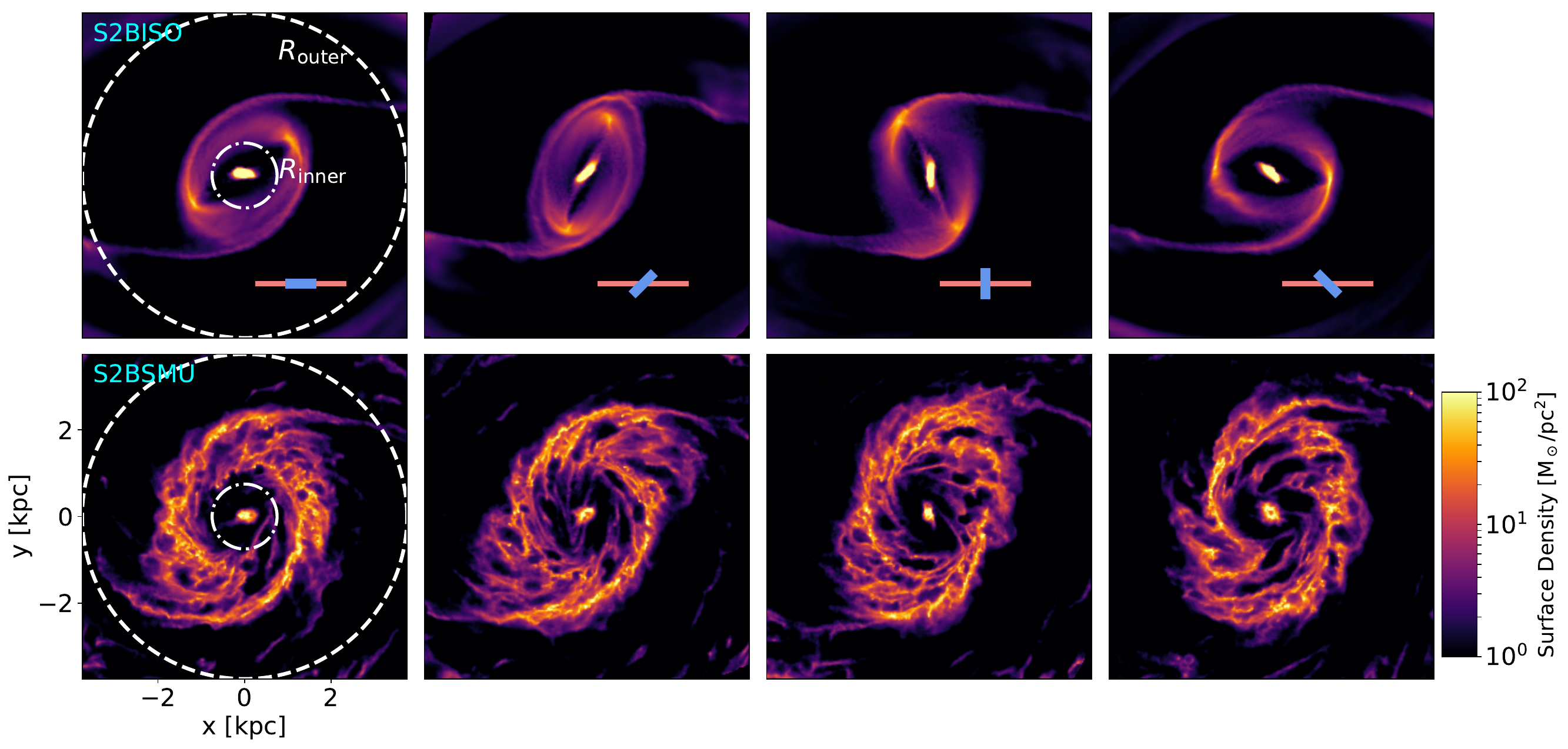}
\caption{Comparison of the inner gas surface density between S2BISO (upper) and S2BSMU (lower). The four columns represent 4 different epochs when the two bars have different angles with respect to each other (indicated by the red and blue sticks at the bottom-right corners of the upper row). The snapshots are taken at $t=1.271,1.287,1.310,1.334\Gyr$ for S2BISO and $t=1.256,1.272,1.295,1.317\Gyr$ for S2BSMU. The dashed and dotted-dashed circles denote two characteristic cylindrical radii of $R_{\rm outer}=3.75\kpc$ and $R_{\rm inner}=0.75\kpc$. The gas disk rotates counterclockwise.
\label{fig:S2Bdencomp}}
\vspace{0.2cm}
\end{figure*}

%fig5
\begin{figure*}[!t]
\includegraphics[width=1.0\textwidth]{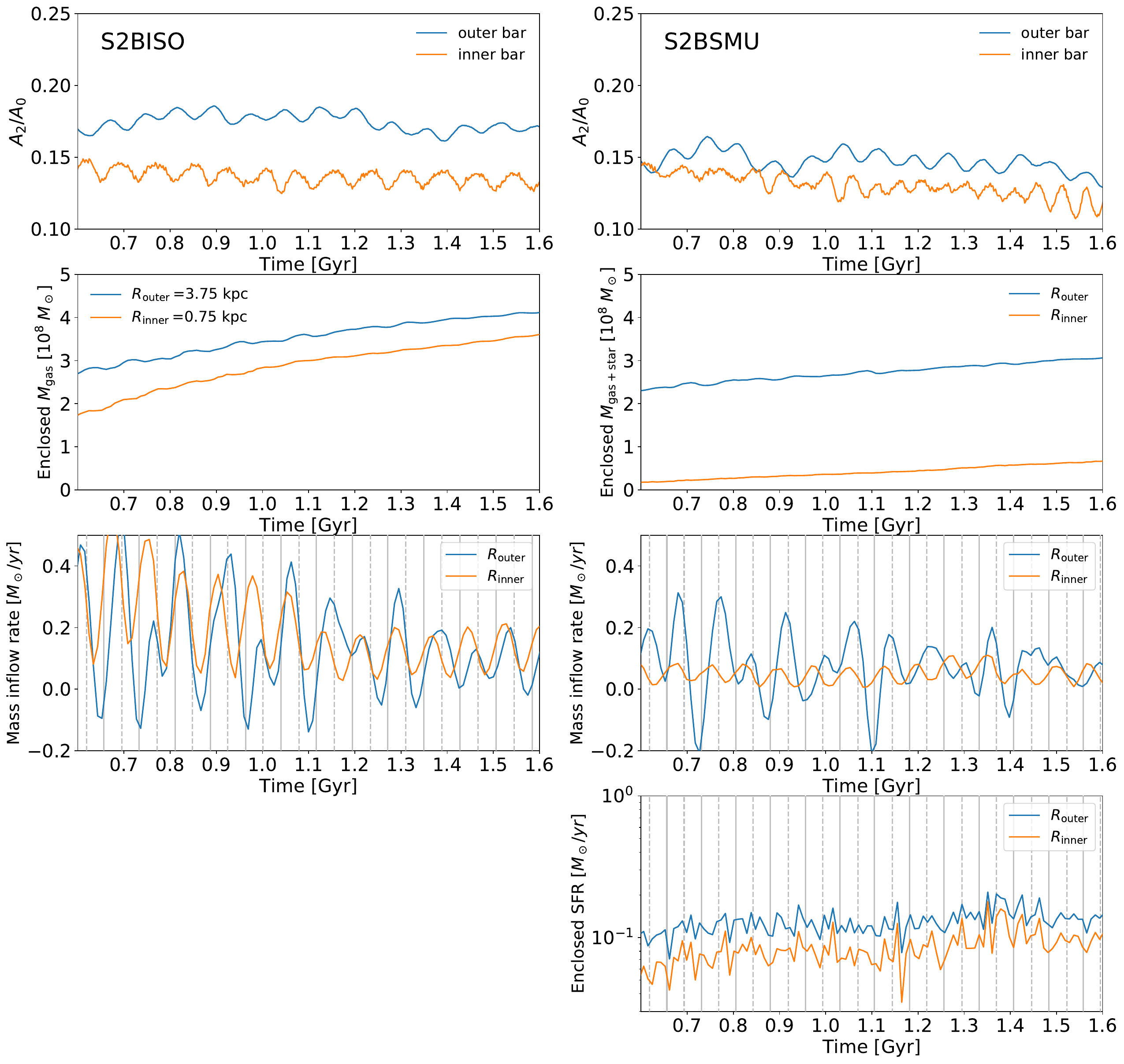}
\caption{Bar properties and gas inflows in S2BISO (left) and S2BSMU (right) models. First row: outer and inner bar amplitudes. The Fourier components ${\rm A_2}$ and ${\rm A_0}$ are measured in the radial range of $2.4-7.5\kpc$ for the outer bar and $0-0.75\kpc$ for the inner bar; Second row: enclosed mass of gas (S2BISO) and gas plus newly formed stars (S2BSMU) inside $R_{\rm inner}$ and $R_{\rm outer}$; Third row: mass inflow rate obtained from the time derivative of the lines in the second row. Vertical grey solid and dashed lines indicate the moments when the two bars are aligned and perpendicular, respectively. Last row: enclosed star formation rate in S2BSMU. This quantity is calculated based on the total mass of new stars inside the given radius that are formed in the past $10\Myr$. 
\label{fig:S2Bisosmuinflow}}
\vspace{0.2cm}
\end{figure*}

%-----------
%-- Sect. 3
%-----------

\section{Results}
\label{sec:result}

\subsection{Comparing S2B Model with Different Gas Assumptions}
\label{sec:isothmulti}

In this subsection, we show how stellar feedback affects the gas morphology as well as the mass inflow rate in the same S2B potential by comparing the fully star-forming model S2BSMU with the isothermal model S2BISO. 

\subsubsection{Gas Morphology in the Inner and Outer Region}
\label{sec:isothmultimorpho}

We first present gas flow patterns in S2BISO with the isothermal EoS. This EoS assumes the specific internal energy of the ISM is the result of a balance between heating and cooling, and can be regarded as a first-order approximation to the observed cold gas in galactic disks. Although it is a simplification of the true multiphase nature of the ISM, previous studies have shown that the isothermal EoS reproduces the observed gas dynamics in bars reasonably well \citep[e.g. ][]{wei_sel_01,lin_etal_13,fragko_etal_17}. We therefore start with this simple model to highlight the gravitational effects of bars. It also helps us to better compare our model with previous S2B simulations that use the isothermal EoS \citep[e.g. ][]{maciej_etal_02,shl_hel_02,nameka_etal_09}.

Figure~\ref{fig:S2Bisoden} plots the surface density of the stellar (top) and the gas (bottom) disk in S2BISO at 4 different times. The models are rotated such that the outer bar is always aligned with the $x$-axis for easier comparison. The two stellar bars are clearly outlined by the white  contours of stellar surface density. Note the inner bar has different orientations with respect to the outer bar due to its different (higher) pattern speed. The first two columns show the earlier evolution of the model. The initial circular and uniform gas disk is gradually distorted by the bars, forming a pair of strong shocks at the leading side of the outer bar together with a two-arm spiral around $10\kpc$. Gas flows inward to the inner boundary soon after the simulation starts (first column), then accumulates at a high-density nuclear ring with a size of $\sim1.5\kpc$ (second column). The inner bar lies within the gaseous nuclear ring, as seen also in observations \citep[][]{buta_etal_15}. Overall, the large-scale flow pattern at $R\gtrsim4\kpc$ resembles the results of previous single-barred simulations \citep[e.g. ][]{kim_etal_12a,li_etal_15,sorman_etal_15a}. The morphology and kinematics of gas in this region is mainly dominated by the outer bar potential, and is not sensitive to the inner bar orientation, as shown in the third (parallel bars) and fourth (perpendicular bars) columns. %The central gas distribution will be discussed later. 

%fig6
\begin{figure*}[!t]
\includegraphics[width=1.0\textwidth]{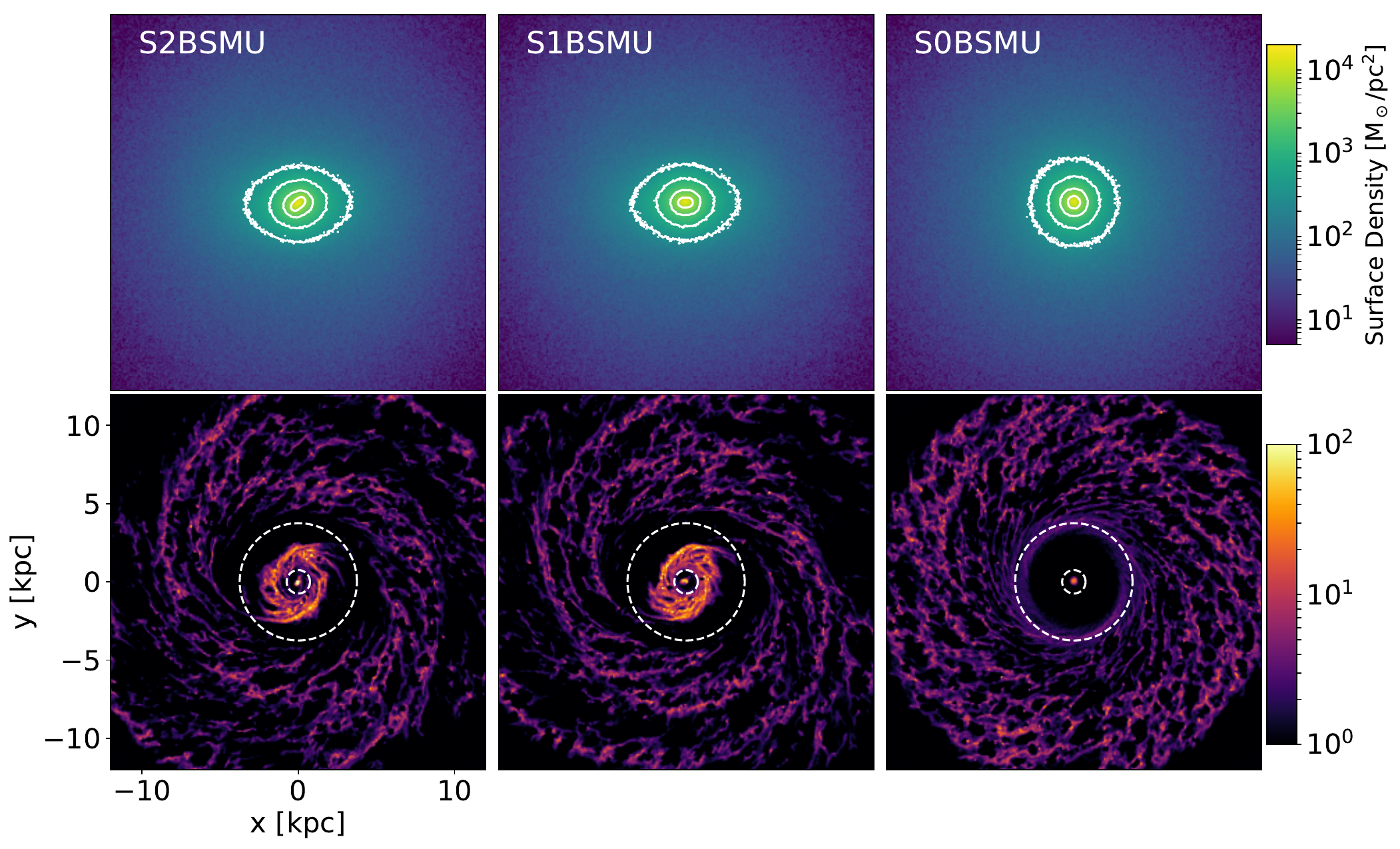}
\caption{Stellar and gas surface density comparison using S2BSMU, S1BSMU, and S0BSMU at $t=1.8\Gyr$, from left to right as labelled. The top row is the stellar surface density and the bottom row is the gas surface density. The outer bar is aligned with the $x$-axis for S2BSMU and S1BSMU. The dashed circles in the bottom row denote the two characteristic cylindrical radii $R_{\rm outer}=3.75\kpc$ and $R_{\rm inner}=0.75\kpc$.
\label{fig:SBsdencomp}}
\vspace{0.2cm}
\end{figure*}

We next present S2BSMU, which resolves the multiphase gas with explicit star formation and stellar feedback included. When comparing the surface densities in Figure~\ref{fig:S2Bsmuden} with those in Figure~\ref{fig:S2Bisoden}, the stellar component is \textit{almost} the same. The smooth gas features (e.g. the shocks, spirals, and the nuclear ring) in S2BISO become more flocculent and clumpy in S2BSMU due to local stellar feedback, but their overall shapes remain mostly unchanged. This is expected since the gas mass fraction is only $2.5\%$ in our models. The global star formation rate (SFR) is around $0.1-0.2\sfrunit$ during the entire simulation, so the gas evolution on the large scale is still governed by the non-axisymmetric stellar potential.

The major differences between S2BISO and S2BSMU emerge in the central $R\lesssim2\kpc$, which is shown in Figure~\ref{fig:S2Bdencomp}. In this figure, we present gas surface densities in 4 snapshots with different bar orientations indicated by the red (outer bar) and blue (inner bar) sticks at the lower right of the first row. The shape of the nuclear ring in S2BISO clearly depends on the relative orientations between the two bars: it tends to be more elliptical when the bars turn from parallel to perpendicular (second and third columns), and vice versa. Furthermore, the ring slightly leads the inner bar during the rotation. The cyclically changing shape of the nuclear ring is consistent with earlier S2B simulations and orbit analysis \citep[][]{mac_spa_00,laine_etal_02}. We also see straight shocks (which would be traced by dust lanes) at the edges of the inner bar (more obvious in the second and third columns) that can drive gas to the center. This flow pattern is different from those found by \citet{maciej_etal_02} and \citet{shl_hel_02}, and we discuss the possible reasons in \S\ref{sec:comparisonwithsims}. On the other hand, the ring in S2BSMU roughly maintains the same shape regardless of the bar relative orientations. This can be understood in terms of a simple estimation of the value of the gas kinetic energy and the energy injected by stellar feedback: the gas in the ring has a typical rotation velocity of $\sim200\kms$ and the gas mass of the ring is $\sim2.0\times10^8\Msun$, which gives a kinetic energy of $8.0\times10^{55}\erg$; the energy released by the supernova (SN) feedback in the ring during one ring rotation ($\sim60\Myr$) is $\approx 6-9\times10^{55}\erg$. The two comparable energies suggest feedback becomes as important as gravity in about one dynamical timescale, and the effect of the inner bar on the ring shape is therefore less prominent. We also find that a small gas bar is formed and co-rotates with the inner bar in S2BISO, but this structure seems to be erased by stellar feedback in S2BSMU. 

%fig7
\begin{figure*}[!t]
\includegraphics[width=1.0\textwidth]{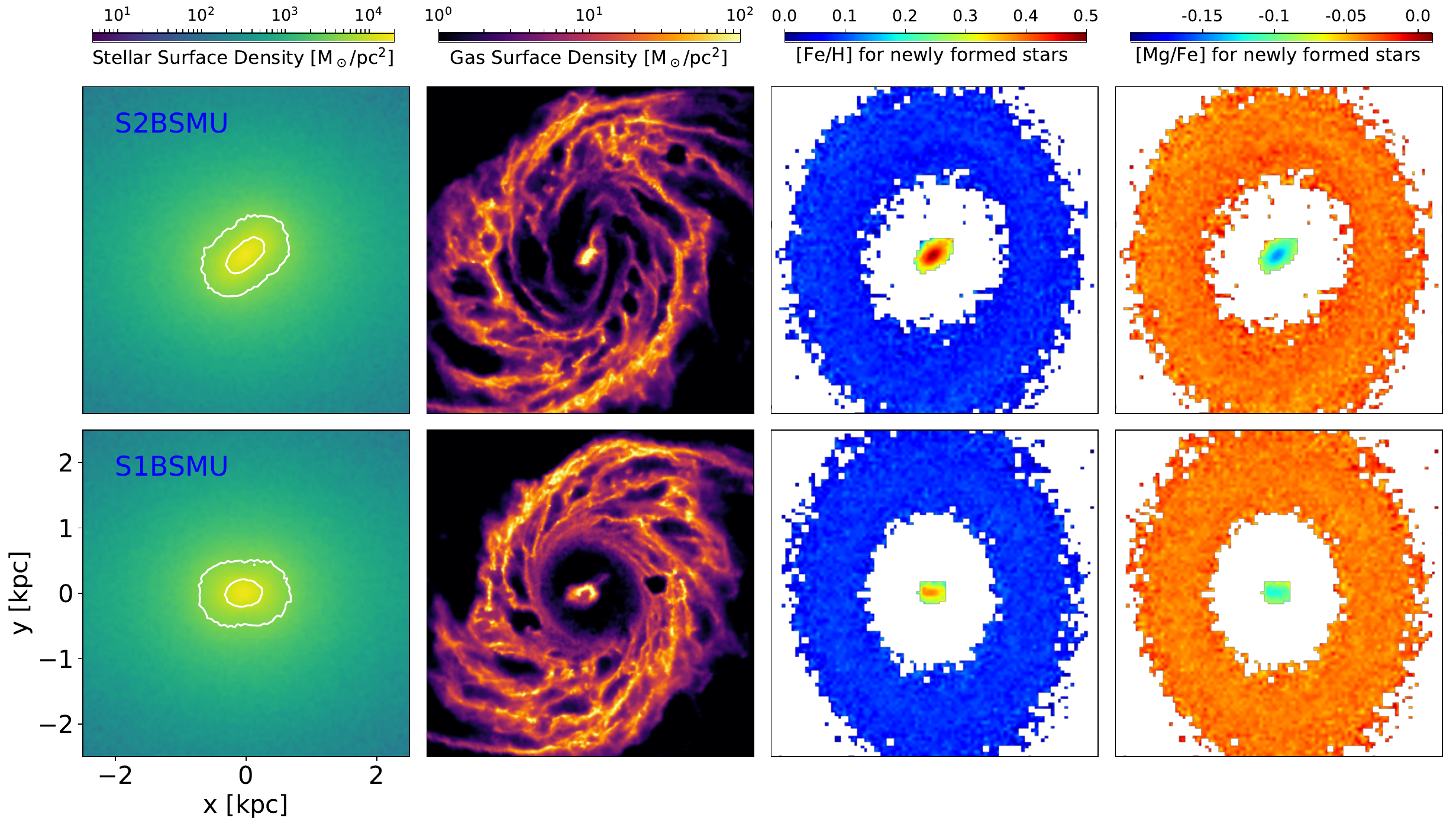}
\caption{A zoom-in view of Figure~\ref{fig:SBsdencomp} to the central $2.5\kpc$. The first and second columns are the stellar and gas surface densities. The third and fourth column plot the metallicity (represented by [Fe/H]) and $\alpha$ abundance (represented by [Mg/Fe]) of the newly formed stars. 
\label{fig:S2BS1Bden}}
\vspace{0.2cm}
\end{figure*}

\subsubsection{Bar Amplitudes and Mass Inflow Rate}
\label{sec:isothmultimdot}

A more quantitative analysis of the evolution of the two S2B models is presented in Figure~\ref{fig:S2Bisosmuinflow}. The two columns display the properties of the isothermal (S2BISO) and the multiphase (S2BSMU) models. We first compare the bar amplitude measured by the normalized Fourier component ${\rm A_2/A_0}$ in the top row. The bars in S2BISO are relatively stable, although the bar amplitudes pulsate as the inner bar rotates with respect to the outer one, which is also seen in previous simulations \citep[][]{deb_she_07,woznia_15,du_etal_15,wu_etal_16}. On the other hand, the bar amplitudes exhibit a weakening trend in S2BSMU. This is because gas self-gravity is included in this model, and the accumulating gas forms a CMC that can weaken the bars \citep[][]{athana_03,she_sel_04,du_etal_17}. In the second row, we plot the enclosed gas mass at two characteristic radii ($R_{\rm outer}=R_1=3.75\kpc$ and $R_{\rm inner}=0.75\kpc$), which can be seen by the white dashed and dotted-dashed circles in the first column of Figure~\ref{fig:S2Bdencomp}. $R_{\rm outer}$ is equal to $R_1$, which corresponds to the inner boundary of the initial gas disk (see \S\ref{sec:sims}), and most of the gas inflow driven by the outer bar accumulates inside this radius; $R_{\rm inner}$ is approximately the size of the inner bar. We see that the enclosed mass quickly increases with time: it reaches $\sim3.0\times10^8\Msun$ inside $R_{\rm outer}$ within $1\Gyr$ in both models. This value is about $20\%$ of our initial gas disk mass and is about $0.5\%$ of the stellar disk mass. Note that the enclosed mass includes both gas and newly formed stars in S2BSMU. It is also worth noting that when the inner bar in S2BSMU starts to weaken at $t\gtrsim1.3\Gyr$, the mass enclosed within $R_{\rm inner}$ reaches $\sim0.1\%$ of the disk mass. This is consistent with the inner bar dissolution criterion predicted in \citet{du_etal_17}. The outer bar in this model is also weakened over time. 

About $80\%$ of the gas in the nuclear ring is accumulated inside $R_{\rm inner}$ in S2BISO, but the fraction drops to $\sim30\%$ in S2BSMU. There are probably two reasons for this: (1) Rapid star formation consumes gas in the nuclear ring at $R\sim2\kpc$ of S2BSMU, which makes less gas available for the inner bar to drive inwards to $R_{\rm inner}$. The gravitational dynamical timescale (i.e. free-fall timescale) for a gas cell to form stars in the ring with a typical gas density of $0.5\dunits$ is around $10\Myr$, while the gas rotation period in the ring is $\sim60\Myr$, and it usually takes a few rotation periods for the inner bar to remove gas angular momentum. Since the dynamical timescale is longer than the local star formation timescale, a large fraction of gas will form stars in the ring instead of being driven inwards by the inner bar. By contrast, there is no mechanism to consume gas in S2BISO, and the inner bar can therefore drive most gas in the ring down to the center. (2) The ring in S2BISO is smaller in size, and periodically changes its shape (Figure~\ref{fig:S2Bdencomp}). This helps gas in the ring to be more easily affected by the inner bar torque in S2BISO compared to S2BSMU.

%fig8
\begin{figure*}[!t]
\includegraphics[width=1.0\textwidth]{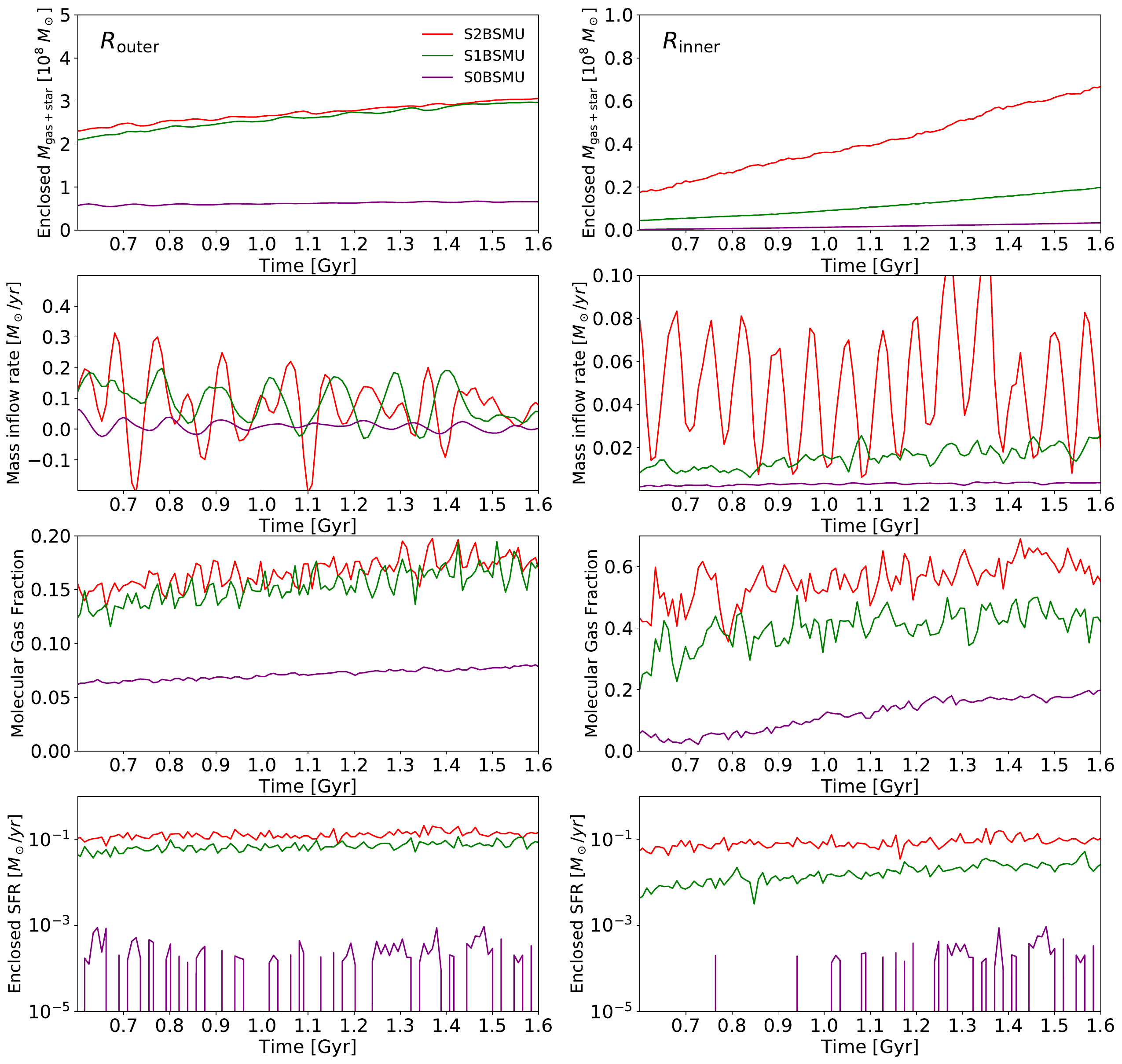}
\caption{Comparison of the enclosed mass (top row), mass inflow rate (second row), molecular gas fraction (third row), and SFR (bottom row) at two characteristic radii ($R_{\rm outer}=3.75\kpc$ and $R_{\rm inner}=0.75\kpc$) of models S2BSMU, S1BSMU, and S0BSMU. The molecular gas fraction is defined as the ratio of molecular gas mass and total gas mass. Note the range of the $y$-axis is not the same in the left and right columns.
\label{fig:SBssfrinflow}}
\vspace{0.2cm}
\end{figure*}

The third row of Figure~\ref{fig:S2Bisosmuinflow} shows the mass inflow rate at the two radii, obtained by taking the time derivative of the lines in the second row. We focus on $0.6-1.6\Gyr$ when the inner bars are relatively stable in both models. The mass inflow rates oscillate with time and seem to be quasi-periodic. We label the moments when the inner bar and outer bar are parallel (solid vertical lines) and perpendicular (dashed vertical lines) to each other. The inflow rate tends to peak when the bars are perpendicular, consistent with more clearly shock features in the top middle two columns of Figure~\ref{fig:S2Bdencomp}. The oscillation period ($\sim78\Myr$) is therefore determined by the pattern speed difference (or the beat frequency) of the two bars. This bar-modulated periodic gas inflow is also reported in some of the previous S2B simulations \citep[][]{shl_hel_02,nameka_etal_09}, but here we show for the first time that the inflow is strongest when the two bars are perpendicular, possibly due to more clear shocks inside the inner bar at this angle. Interestingly, this trend remains in S2BSMU, although the spatial distribution of gas seems to be independent with respect to the bar orientations due to local stellar feedback (see the bottom row in Figure~\ref{fig:S2Bdencomp}). It is worth noting that there appears to be no regular pattern in the enclosed SFR curve of S2BSMU within $R_{\rm inner}$, despite the fact that the model has a nearly periodic gas supply in this region \citep[but also see the results in ][]{moon_etal_22}.

Figure~\ref{fig:S2Bisosmuinflow} demonstrates that the inner bar \textit{does} promote gas inflow to the center. However, the mass inflow rate at $R_{\rm inner}$ in S2BSMU is considerably lower than that in S2BISO. It is therefore natural to ask whether the amplitude of the central mass inflow rate is still dominated by the inner bar potential when stellar feedback is included.

%fig9
\begin{figure}[!t]
\includegraphics[width=0.5\textwidth]{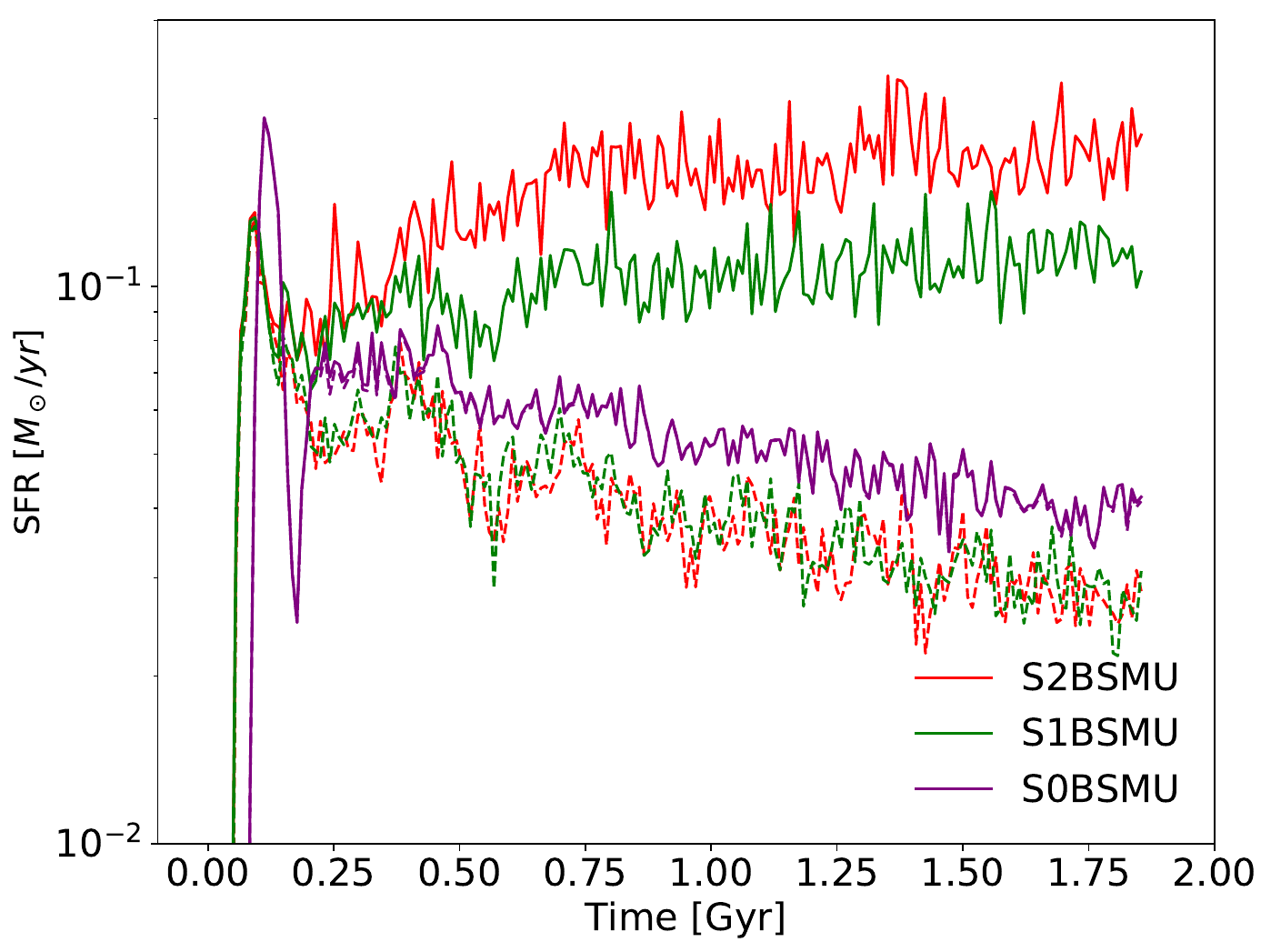}
\caption{Solid lines: total SFR as a function of time in S2BSMU, S1BSMU, and S0BSMU. Dashed lines: SFR at $R\geq R_{\rm outer}=3.75\kpc$ as a function of time in the models.
\label{fig:SBssfrall}}
\vspace{0.2cm}
\end{figure}

\subsection{Comparing S2B Model with SB and SA Models}
\label{sec:barnumber}

In this subsection we compare three models with different bar numbers but all with the {\tt SMUGGLE} model (S2BSMU, S1BSMU, S0BSMU) to isolate the effects of the inner bar on gas dynamics from (stochastic) stellar feedback. 

\subsubsection{Mass Inflow Rate and Star Formation Rate}
\label{sec:sbsmdot}

%fig10
\begin{figure*}[!t]
\includegraphics[width=1.0\textwidth]{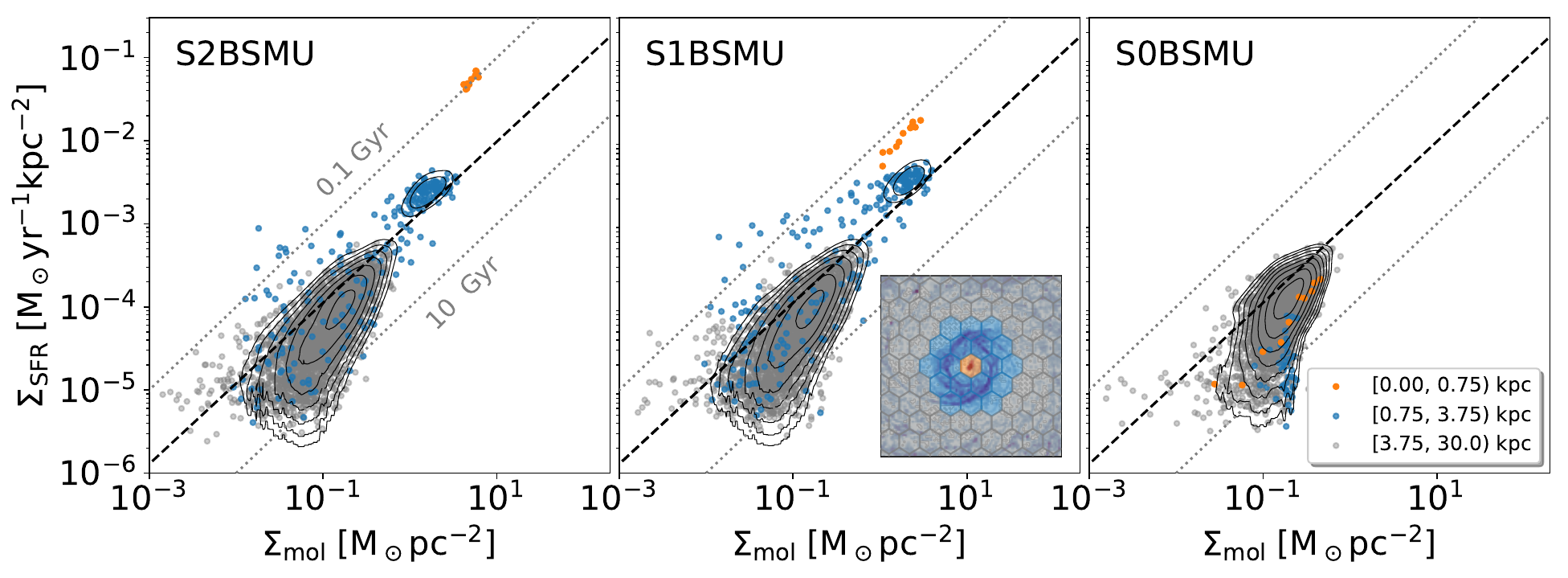}
\caption{KS relation in models S2BSMU, S1BSMU, and S0BSMU. Solid black curves are the contours of the scatter points. Only gas and newly formed stars in the disk region ($\abs{z}\leq2\kpc$) are considered in this plot. For each model, we include 11 snapshots separated by $\sim100\Myr$ to increase the sample size. The black dashed line is the global fit from \citet{querej_etal_21} with a slope of $0.97$. Two black dotted lines representing the depletion time of $0.1\Gyr$ and $10\Gyr$ are included for comparison. The subplot in the middle panel illustrates the placement of the hexagonal bins in our models. The bins are colored according to their galactocentric radius, and the underlying plot is the same gas surface density plot of S2BSMU shown in Figure~\ref{fig:S2BS1Bden}. 
\label{fig:SBsKSplot}}
\vspace{0.2cm}
\end{figure*}

Figure~\ref{fig:SBsdencomp} shows the stellar and gas surface densities of S2BSMU, S1BSMU, and S0BSMU at $t=1.8\Gyr$. These three models have the same azimuthally-averaged mass distribution at $t=0$. Gas flow patterns in S2BSMU and S1BSMU are very similar outside $R\sim1\kpc$; they both form a pair of spiral arms and a bright nuclear ring produced by the outer bar. S0BSMU only shows flocculent spirals and little gas can flow into $R_{\rm outer}$ as no bar is present\footnote{The faint gas blob at the center of S0BSMU is probably due to numerical diffusion.}. The gas distribution of this model looks similar to NGC~2775 \citep[e.g. ][]{leroy_etal_21}. 

Inside $R\sim1\kpc$, the gas behaves differently in S2BSMU and S1BSMU as shown in Figure~\ref{fig:S2BS1Bden}. The left two columns are zoom-in views of the stellar and gas surface densities in Figure~\ref{fig:SBsdencomp}. Inflow signatures can be spotted in S2BSMU as multiple gas streams formed inside the nuclear ring down to the galactic center. On the other hand, the nuclear ring has a relatively well-defined inner boundary in S1BSMU, implying the inflow has stalled at this radius. This is not surprising since the nuclear ring in single-barred galaxies serves as a barrier that prevents gas from reaching the nucleus, even if stellar feedback is included \citep[][]{fanali_etal_15,li_etal_17,tress_etal_20}. The right two columns illustrate the spatial distribution of the newly formed stars color-coded by their chemistries ([Fe/H] and [Mg/Fe]). Although we do not observe an inner gas bar in S2BSMU, newly formed stars inside the ring follow the shape and rotation of the stellar inner bar, similar to the cascade bar formation scenario proposed in previous studies \citep[e.g. ][]{shlosm_etal_89,shlosm_etal_90,hop_qua_10}. This bar-shaped, newly formed stellar structure is more metal rich and less $\alpha$-enhanced compared to the surroundings, consistent with the observations of the inner bar in NGC~5850 \citep{lorenz_etal_19b}. The newly formed stars in S1BSMU show a similar pattern, but not as strongly as in S2BSMU.

In Figure~\ref{fig:SBssfrinflow} we plot the enclosed mass, mass inflow rate, and SFR for these three models at $R_{\rm outer}$ (left column) and $R_{\rm inner}$ (right column). The accumulated masses at $R_{\rm outer}$ are almost identical for S2BSMU and S1BSMU, but the two models diverge inside $R_{\rm inner}$. The inner bar in S2BSMU drives $\sim3$ times more mass inside $R_{\rm inner}$ compared with the single-barred model. The mass inflow rate of S2BSMU at $R_{\rm inner}$ is quasi-periodic and is related to the inner bar dynamics, as shown in Figure~\ref{fig:S2Bisosmuinflow}. By contrast, the mass inflow rate of S1BSMU at $R_{\rm inner}$ is mostly dominated by stellar feedback, and the periodicity disappears. We obtained a time-averaged mass inflow rate of $0.144\sfrunit$ at $R_{\rm outer}$ and $0.013\sfrunit$ at $R_{\rm inner}$ for S1BSMU, within the time interval of $t=0.6-1.6\Gyr$; the ratio of these two inflow rates is $\sim10.8$, similar to the results in \citet{tress_etal_20}. S2BSMU increases the central inflow rate to $0.044\sfrunit$, reducing the ratio to $\sim3.4$. The control model S0BSMU has weak inflows (probably caused by diffusion and numerical effects), which are negligible compared with the other two models. We therefore conclude that the inner bar in S2Bs indeed helps gas flowing inwards. 

%fig11
\begin{figure}[!t]
\includegraphics[width=0.45\textwidth]{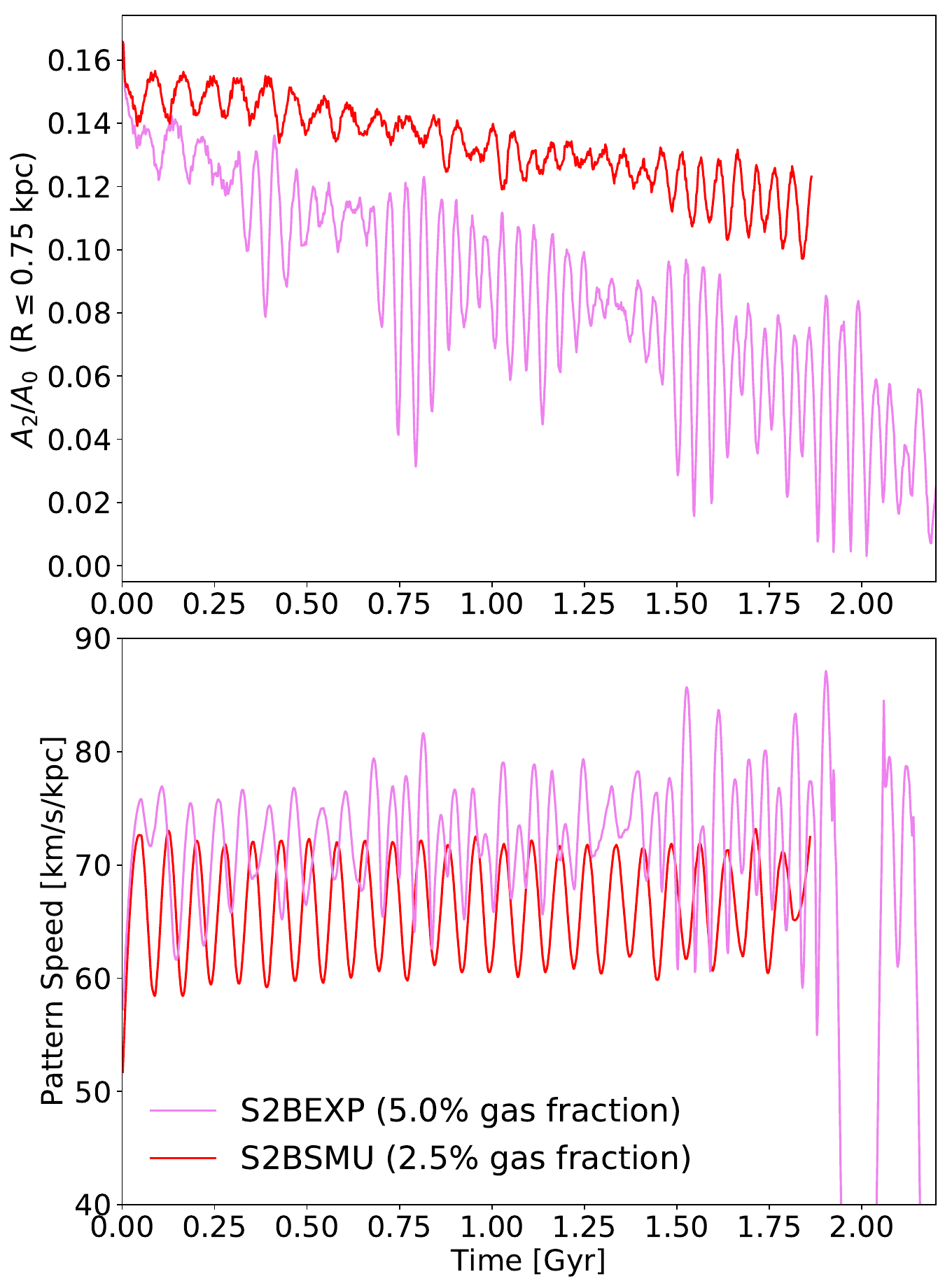}
\caption{Evolution of ${\rm A_2/A_0}$ (top) and pattern speed (bottom) in the inner bar region ($R\leq0.75\kpc$) of two models with different gas fraction. At $t\gtrsim1.7\Gyr$ in S2BEXP, the pattern speed calculated using the ${\rm A_2}$ phase angle time derivatives may be less accurate due to the small ${\rm A_2}$ amplitude.
\label{fig:innerbardiss}}
\vspace{0.2cm}
\end{figure}

%fig12
\begin{figure*}[!t]
\includegraphics[width=1.0\textwidth]{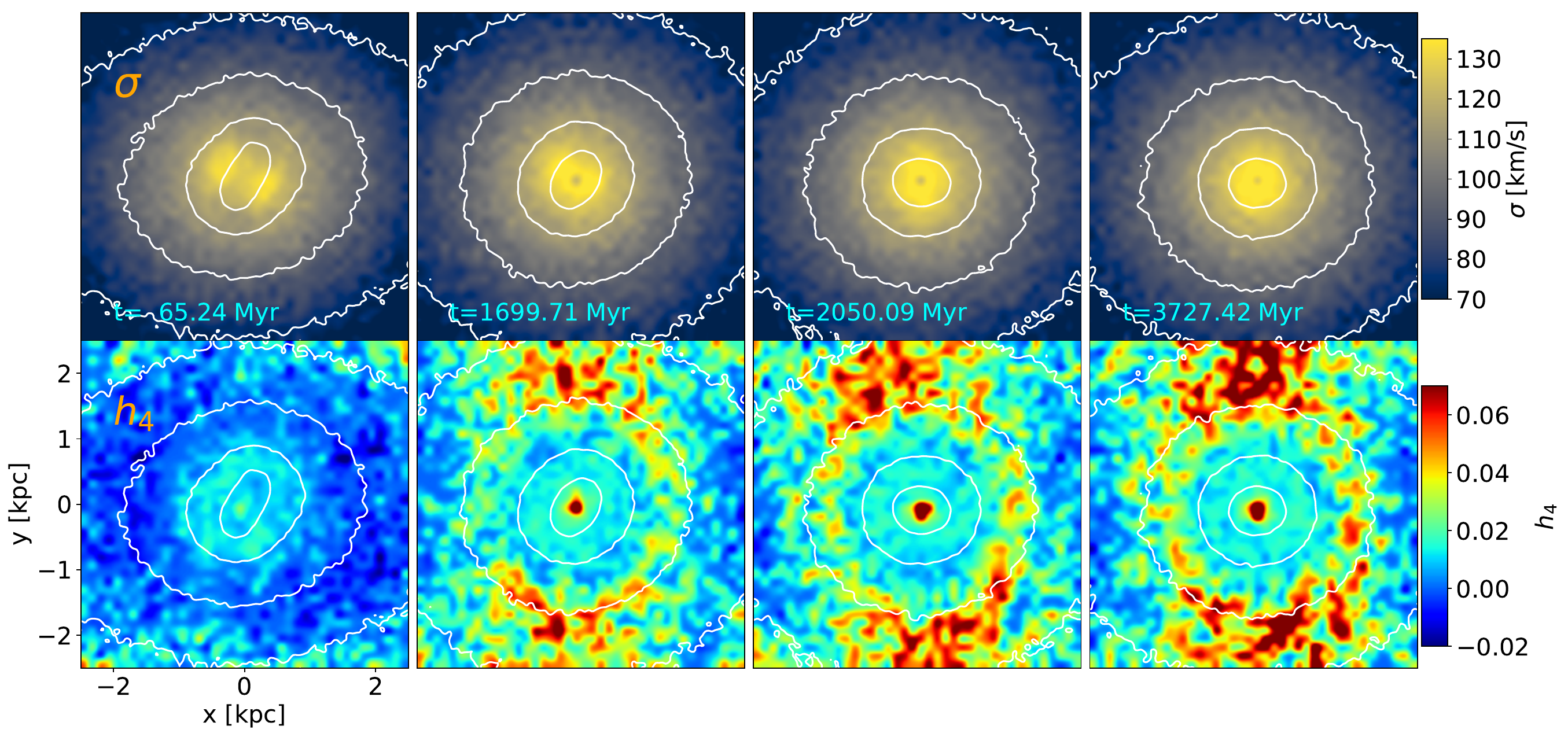}
\caption{Kinematic maps of S2BEXP, viewed face-on. From left to right are 4 different epochs. The second ($\sigma$) and fourth ($h_4$) Gauss–Hermite moments of the vertical velocity $v_z$ distribution are shown in the top and bottom rows, respectively. White lines represent the contours of the stellar surface density. The outer bar is aligned with the $x$-axis. 
\label{fig:sigmavz}}
\vspace{0.2cm}
\end{figure*}

In addition, we find S2BSMU has the highest SFR in both regions as shown in the last row of Figure~\ref{fig:SBssfrinflow}. This is expected as a consequence of more accumulated gas in the center. Observations have shown that bars can enhance central star formation \citep[e.g. ][]{lin_etal_17,lin_etal_20}, and our results imply multiple bars may further amplify this effect. In fact, the total SFR in S2BSMU is larger than the other two models during most of the time in our simulations (solid lines in Figure~\ref{fig:SBssfrall}). This indicates that bars may not significantly reduce the SFR within a short period in our gas-poor models \citep[but see also][]{khoper_etal_18}. The SFR in S2BSMU and S1BSMU at $R>R_{\rm outer}$ is slightly lower than that in S0BSMU (dashed lines in Figure~\ref{fig:SBssfrall}). This is probably due to a smaller gas surface density in the outer disk region, as a significant fraction of gas has been driven inside $R_{\rm outer}$. 

\subsubsection{Kennicutt–Schmidt Relation}
\label{sec:sbskslaw}

We investigate also the molecular Kennicutt–Schmidt (KS) relation \citep{schmid_59,kennic_98} that connects the molecular gas surface density ($\Sigma_{\rm mol}$) to the star formation rate surface density ($\Sigma_{\rm SFR}$) in our models. Motivated by \citet{querej_etal_21} who used PHANGS-ALMA data \citep{leroy_etal_21} to explore the KS relation in different galactic environments, we dissect our simulations into $\sim1.3\kpc$-wide hexagonal bins\footnote{See the subplot in the middle panel of Figure~\ref{fig:SBsKSplot}, which is similar to Fig.~3 in \citet{querej_etal_21}.}. The gas and SFR surface densities are then estimated inside each bin and are plotted as colored dots in Figure~\ref{fig:SBsKSplot}. We divided the bins into three groups based on their radius: inside the inner bar (orange); outside the inner bar but roughly covering the nuclear ring (blue); and the disk region (grey). In general, the KS relation in our model is similar to the global fit in \citet{querej_etal_21} (black dashed line), which implies a weak environmental (or radial) dependence of the star formation efficiency. Centers (orange points) in barred models (S2BSMU and S1BSMU) have shorter depletion times than the rest of the disk, also consistent with the findings in \citet{querej_etal_21}. In addition, the depletion time in the center of S2BSMU seems to be even smaller than that in S1BSMU. This is possibly due to a higher molecular gas fraction in the central regions of S2BSMU, as shown in the third row of Figure~\ref{fig:SBssfrinflow}. On the other hand, the low gas surface density as well as SFR in the center of the unbarred model S0BSMU is simply due to our centrally-holed initial gas disk setup, and thus cannot reflect the real central environments in unbarred galaxies. We therefore conclude that {\tt SMUGGLE} model can reproduce the latest observations reasonably well. In {\tt SMUGGLE} model we adopt a constant star formation efficiency per free-fall time ($\epsilon_{\rm sf}=0.01$) and compute the SFR based on the molecular gas only, the KS relation is then captured by the self-regulated star formation and feedback processes. However, there are hints that $\epsilon_{\rm sf}$ may vary systematically with $\Sigma_{\rm mol}$ and $\Sigma_{\rm SFR}$ \citep{sun_etal_23}. We leave a detailed comparison of the star formation laws between simulations and observations for future studies. 

%fig13
\begin{figure*}[!t]
\includegraphics[width=1.0\textwidth]{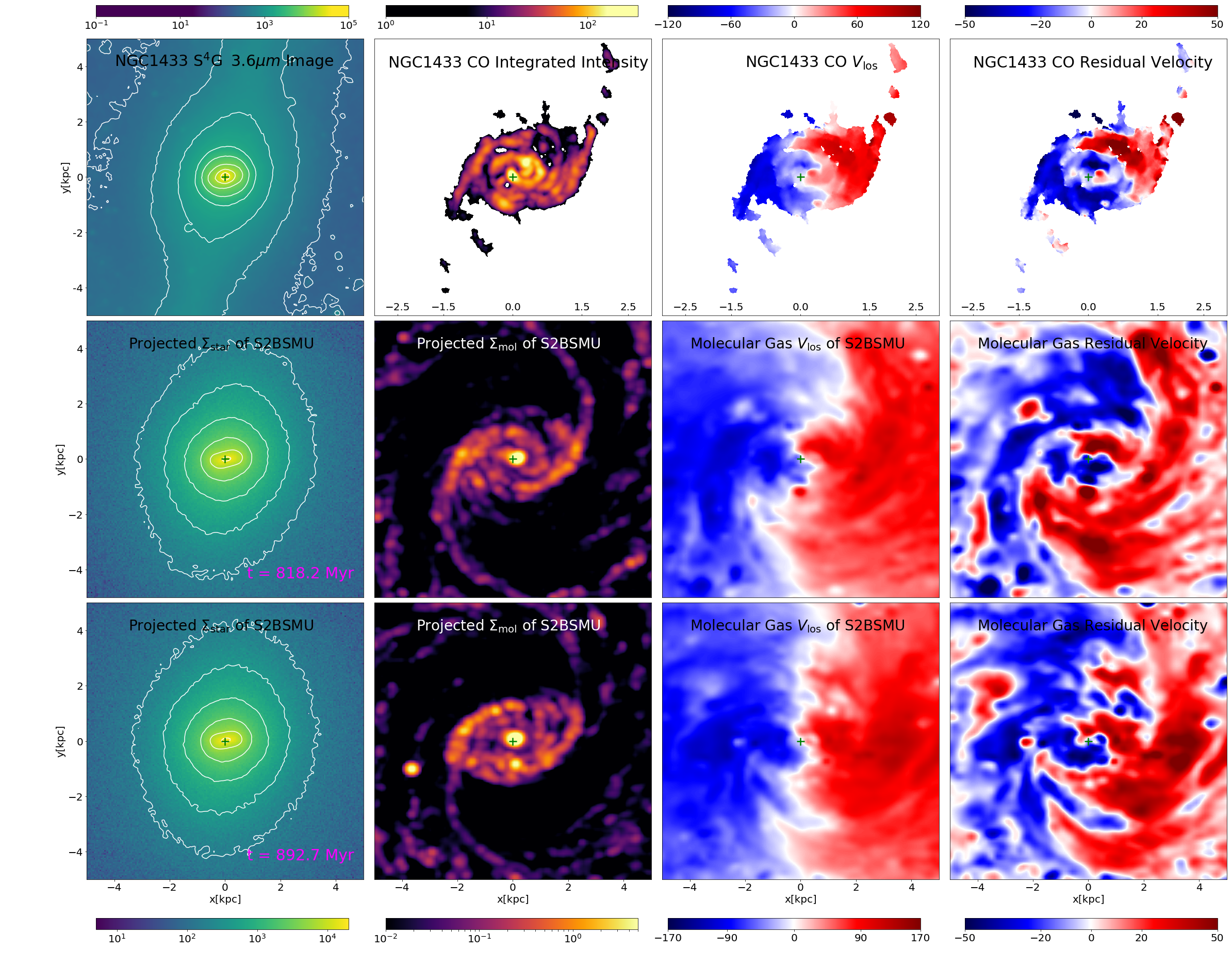}
\caption{Comparison between an observed S2B galaxy NGC~1433 and our S2BSMU model. We adopt a distance of $18.63\Mpc$, a systemic velocity of $1057.4\kms$, a disk position angle of $199.7\degree$, and an inclination angle of $28.6\degree$ for this galaxy \citep[][]{buta_etal_01,lang_etal_20}. Top row: from left to right are S$^{4}$G $3.6\;\mu m$ image (unit: ${\rm L_\odot/pc^2}$), PHANGS-ALMA CO(2-1) integrated intensity map (moment 0, unit: ${\rm K\kms}$), intensity weighted mean velocity map (moment 1, unit: ${\rm km\;s^{-1}}$), and residual velocity map obtained by subtracting circular motions (unit: ${\rm km\;s^{-1}}$). Note the spatial scales are different for S$^{4}$G image and ALMA maps. Second row: from left to right are projected stellar surface density map (unit: ${\rm M_\odot/pc^2}$), projected molecular gas surface density map (unit: ${\rm M_\odot/pc^2}$), molecular gas line-of-sight velocity map (unit: ${\rm km\;s^{-1}}$), and molecular gas residual velocity map (unit: ${\rm km\;s^{-1}}$) for the model at $t=818.2\Myr$. Bottom row: same as the second row but at $t=892.7\Myr$. The green plus sign marks the center of the image $(x=0,y=0)$. 
\label{fig:ngc1433}}
\vspace{0.2cm}
\end{figure*}

\subsection{Long Term Evolution}

Figure~\ref{fig:S2Bisosmuinflow} has shown that the inner bar in S2BSMU gradually decays due to accumulated central mass. We therefore use S2BEXP to test how long the inner bar will survive if there is an even higher gas mass at the center from the beginning. The gas mass inside $R_{\rm inner}$ is $\sim0.1\%$ stellar mass at $t=0$ for this model, which meets the inner bar dissolution criterion in \citet{du_etal_17}. The evolution of the Fourier component {\rm $A_2/A_0$} and inner bar pattern speed is shown in Figure~\ref{fig:innerbardiss}. We find that the independent pattern speed inside $R_{\rm inner}$ disappears at $t\sim2\Gyr$. At this moment, the contours of the stellar surface density in the central $1\kpc$ are nearly round (see the third column of Figure~\ref{fig:sigmavz}). We conclude the inner bar fully dissolves at $t\sim2\Gyr$ in S2BEXP. The second and third columns of Figure~\ref{fig:sigmavz} are two snapshots before and after the dissolution of the inner bar. We note the outer bar also weakens considerably in this model.   

We plot the evolution of 2D stellar kinematic maps of S2BEXP in Figure~\ref{fig:sigmavz}. The top and bottom row shows the second ($\sigma$) and fourth ($h_4$) Gauss–Hermite moments \citep[][]{gerhar_93,van_fra_93} of the vertical velocity ($v_z$) distribution. We include both the old stars and the newly formed stars in this figure. The $\sigma$-humps \citep[][]{lorenz_etal_08,du_etal_16} along the minor axis of the inner bar is clearly seen in the first column, but this feature disappears when the inner bar has fully been destroyed. The central region then remains roughly axisymmetric till the end of the simulation \citep[see also Fig.~8 in ][]{guo_etal_20}. The $h_4$ maps reveal a ring-like structure and a central blob that peak better after $t\sim1.7\Gyr$, which are possibly associated with the newly formed stars in the simulation. Even though the star formation is high in the inner bar region, the newly formed stars mainly follow chaotic orbits scattered by the CMC, and can hardly support the inner bar structure as suggested by \citet{nak_bab_23}. Strong stellar feedback of young stars may also partially contribute to the dissolution of the inner bar, as it prevents star forming clouds settling on the inner bar supporting orbits (see also \S\ref{sec:isothmultimorpho} and Figure~\ref{fig:ngc1433}). These kinematic signatures could be potentially useful for detecting dissolved inner bars in future integral field unit (IFU) observations.

%-----------
%-- Sect. 4
%-----------

\section{Comparison with Observations}
\label{sec:compobs}

Inflow signatures related to the inner bar have been reported in observations \citep[e.g. ][]{fathi_etal_06,schinn_etal_06,schinn_etal_07,gonzal_etal_21}. Based on the recent PHANGS-ALMA data, we find 6 S2B candidates (NGC~1068, NGC~1087, NGC~1317, NGC~1433, NGC~4304, and NGC~4321) according to the classifications in \citet{erwin_04} and \citet{buta_etal_15}. NGC~1433 probably has the most similar bar shapes (i.e. stellar surface density contours) compared to our S2BSMU model. Although our model S2BSMU is not designed to match any particular galaxy, we still compare it with NGC~1433 in Figure~\ref{fig:ngc1433}. The top row shows the S$^4$G $3.6\;\mu m$ image, CO(2-1) integrated intensity map, CO intensity-weighted mean line-of-slight (LOS) velocity map, and the residual velocity map obtained by subtracting the circular motion (i.e. the spider diagram) from the $V_{\rm los}$ map of NGC~1433. The first two CO maps are the same as in the PHANGS-ALMA survey paper \citep[moment 0 and moment 1 maps in][]{leroy_etal_21}. The circular velocity curve used to derive the residual velocity map is from \citet{lang_etal_20} (see their Eq.~10 and Table~4, with $V_0=204.5\kms$ and $R_t=0.59\kpc$). The middle and bottom rows show the projected stellar surface density map, projected molecular gas surface density map, molecular gas LOS velocity map, and residual velocity map in S2BSMU at $t=818\Myr$ and $892\Myr$. These plots are obtained by projecting the stellar and molecular gas disk in the simulation with an inclination of $28.6^\circ$ and a disk position angle of $199.7^\circ$ \citep[][]{buta_etal_01,lang_etal_20}. The circular velocity curve is derived by {\tt DiskFit} (\citealt{sel_spe_15}) that applies the tilted ring method\footnote{The tilted ring method may fail to extract the true circular velocity curve when bar-induced non-circular motions dominate the LOS velocity field. A more detailed analysis will be presented in Liu et al. (in prep.).} to the molecular gas $V_{\rm los}$ field ($V_0=270.0\kms$ and $R_t=0.30\kpc$). We assume purely circular motion and do not consider radial or bisymmetric flows during the fitting, similar to \citet{lang_etal_20}.

In Figure~\ref{fig:ngc1433}, the two snapshots of S2BSMU have similar bar orientations as in the S$^4$G image (first column). In the second column, multiple gas streams extending from the nuclear ring to the central part are seen both in the observation and in our model. This indicates a strong gas inflow that is also evident in the other five observed S2Bs. The presence of clumpy gas features around the nuclear ring suggests that these regions are undergoing intense star formation. The $V_{\rm los}$ maps in the third column also match qualitatively. Note here we show the velocity information in the whole region of the simulation, which presents a larger view of gas flows. We argue that reproducing exactly the observed CO $V_{\rm los}$ map of S2Bs would be quite challenging, since the effects of (stochastic) stellar feedback on gas morphology and kinematics are as important as the gravitational effect of the inner bar (\S\ref{sec:isothmultimorpho}). We also demonstrated this in the second and third rows of Figure~\ref{fig:ngc1433}: the stellar surface densities are almost identical at $t=818\Myr$ and $892\Myr$, but the gas morphology and $V_{\rm los}$ maps have a relatively large difference. In the fourth column we show the residual velocity maps of the observation and simulation. The residual velocity map in the simulation look more chaotic, possibly also due to strong stellar feedback. We note the structures in the residual velocity map somewhat relies on whether people can measure the inclination and position angle of disks accurately \citep[e.g. ][]{kolcu_etal_23}, thus a large uncertainty may exist to make a reasonable comparison.

Another way to compare our model with real S2Bs is to calculate the mass inflow rate at different radii. We predict that the ratio of the mass inflow rate from the outer bar to the nuclear ring and from the nuclear ring to the center is around 3.4 for S2Bs, while this number increases to $\gtrsim10$ for single-barred galaxies (\S\ref{sec:barnumber}). \citet{wu_etal_21} obtained a mass inflow rate of $12\sfrunit$ along the dust lanes of the outer bar in the S2B galaxy NGC~3504 based on ALMA CO data. It would be interesting to measure the radial motions of gas in this galaxy (as well as other S2Bs) at different radii, and compare the results with single-barred galaxies \citep[e.g. similar to ][]{haan_etal_09}. However, this requires a detailed analysis of the observational data, which is the subject of future works.

%-----------
%-- Sect. 5
%-----------

\section{Discussion}
\label{sec:discussion}

\subsection{Comparison with previous S2B models}
\label{sec:comparisonwithsims}

Our models suggest that inner bars are efficient at driving gas inflows, which does not support the previous S2B simulations in \citet{maciej_etal_02} and \citet{rautia_etal_02}. We first note that the results in \citet{maciej_etal_02} may suffer from a now-recognized bug in the code {\tt CHMOG} \citep[see ][]{kim_etal_12a}. Indeed \citet{nameka_etal_09} adopted a similar analytical S2B potential but a different nuclear bulge component compared with \citet{maciej_etal_02} and found significant inflows caused by the inner bar. The authors speculated that the shallow central potential in \citet{maciej_etal_02} may be the reason for the lack of inflow. While the model of \citet{rautia_etal_02} consisted of a live S2B model, its resolution may have been too low to resolve the central gas behavior since it employed only 40000 sticky gas particles within a $4.5\kpc$ radius gas disk. Thus, it is possible that the absence of strong gas inflows in \citet{maciej_etal_02} and \citet{rautia_etal_02} may mostly be attributed to numerical issues.

The inflow due to the inner bar is quasi-periodic in our S2B models. \citet{shl_hel_02} and \citet{nameka_etal_09} found a similar cyclic inflow in their S2B simulations, although the pattern speeds and the shapes of the bars in their models are constant along time. This may imply a pulsating inner bar is not necessary to produce a periodic inflow. The periodicity is likely due to the fact that the inner bar sweeps gas more efficiently (or shocks are more easily formed) when the two bars tend to be perpendicular with each other, i.e. a pattern speed differences for the two bars would be sufficient. We also find the inner bar in our S2BEXP model does not revive after dissolution, which appears to differ from \citet{woznia_15} who also included star formation and stellar feedback. This may imply that star formation is not the only key parameter for generating a recurrent inner bar, or that the gas fraction is still too low in our models. It could also be due to the different implementations of the sub-grid physics in the two studies.

\subsection{For how long does the inner bar promote gas inflows?}

According to S2BEXP model, the inner bar is fully dissolved within $\sim2\Gyr$ with a gas disk of $\sim5\%$ stellar mass. The gas inflow then stalls at the nuclear ring, similar to the case in S1BSMU. This timescale is longer than that obtained in \citet{du_etal_17} ($\lesssim1\Gyr$), probably because the central potential of the CMC is more `soft' than a nearly Keplerian one used in \citet{du_etal_17}. 

One important aspect that is not considered in our study is the feedback due to the active galactic nucleus (AGN). The huge energy released by the central engine may help to clear the accumulated gas at the center. The inner bar is then expected to be more robust against gas inflows than what we have found here. It is therefore reasonable to suggest that the inner bar can promote central gas inflows on a timescale of a few Gyrs, although the inflows may be episodic. On the other hand, \citet{irodot_etal_22} suggested there is an anti-correlation between AGN feedback and bar strength in cosmological simulations, i.e. bars tend to be stronger and shorter if AGN feedback is not included. Future studies of the relation between AGN feedback and bar stability would be very useful to better understand this issue.

\subsection{Do we expect a clear correlation between S2Bs and AGNs?}

In our S2B model, gas is funneled close to the Bondi radius (a few tens of parsecs) if a $1.0\times10^8\Msun$ SMBH (Schwarzschild radius: $R_S=9.57\times10^{-6}\pc$, Eddington accretion rate: $\dot{M}_{\rm Edd}=2.2\sfrunit$) had been present at the galatic center. The accretion flows around a SMBH are generally classified into the radiative mode, also known as quasar mode, which operates when the SMBH has high accretion rates; and kinetic mode, also known as radio mode, which is typically important when the SMBH has low accretion rates. The boundary between the two modes is usually $\sim2\%\dot{M}_{\rm Edd}=0.043\sfrunit$ \citep[summarized in][]{yuan_etal_20}. Our fiducial S2B model with a lightweight gaseous disk has the same order of gas inflow rate, and this level of fueling can be easily achieved in local disk galaxies that host quasars \citep[][]{zhao_etal_19}. If the SMBHs in S2Bs are in quasar mode, the timescale for the gas to flow from the accretion disk ($R=1000R_S\approx0.01\pc$) to feed the SMBH is $0.44\Gyr$ based on Eq.~6 of \citet{yuan_etal_18}. With an additional free-fall timescale (a few tens of Myr) for gas to flow into the accretion disk from the Bondi radius, the overall timescale of gas inflows from the Bondi radius to feed the central SMBH can be estimated at a few hundred Myr, which is longer than the rotation period of the inner bar in our model. Though gas inflows are more prominent when the inner bar is perpendicular to its outer counterpart, it is not likely to result in a clear correlation between the relative orientation of the two bars and any AGN activity. 

Do we expect to find a higher frequency of AGNs in S2B galaxies? Our simulations suggest that SMBHs may grow rapidly when an inner bar exists. About $10^7\Msun$ of gas accumulates at the central $100\pc$ of S2BSMU within $0.5\Gyr$ even with a rather lightweight initial gaseous disk. As suggested in \citet{du_etal_17} and \citet{guo_etal_20}, and also from the results of this paper, the inner bar will be destroyed once enough gas accumulates in the galaxy center, no matter whether the gas has fallen into the SMBH. The inner bar may be in the process of dissolution, considering a potential delay of AGN activity by around a few hundred Myr estimated above. We also note that any possible link between S2Bs and AGNs may disappear if the first inner bar that powers the AGN decays and a new one forms at later times, similar to the scenario proposed for large-scale bars \citep[e.g. ][]{sel_mor_99,li_etal_17}. It is therefore understandable that no clear relation between inner bars and AGNs is reported in observations \citep[][]{laine_etal_02,erw_spa_02,erwin_11}. On the other hand, recent studies seem to suggest that AGNs are preferentially found in barred galaxies \citep[][]{alonso_etal_18,silval_etal_22}, although the S2B fraction in the samples remains unclear. 

\subsection{Are inner bars long-lived or recurrent?}

The high frequency of observed S2Bs \citep[][]{erwin_11} indicates that inner bars are either long-lived or easily reformed after dissolution.  \citet{lorenz_etal_19a} performed a detailed photometric analysis of 17 nearby S2Bs and found that most bulges in their sample have intermediate S\'{e}rsic index ($n\approx2$) and lay at the top sequence of the Kormendy relation \citep[][]{kormen_77}, i.e. they are similar to classical bulges. On the other hand, the simulation of \citet{guo_etal_20} show that the remnant of the dissolved inner bar share such properties with those of observed bulges in S2Bs. This makes one wonder if the inner bar in S2B can reform after dissolution instead of being a long-lived structure. If inner bars have dissolved and reformed over time, we should see these classical bulges as relics in observed S2Bs, which seems to be consistent with \citet{lorenz_etal_19a}. However, since there are multiple ways to form classical bulges, this does not necessarily mean that inner bars have to be recurrent. It is possible that two kinds of inner bars exist in the universe as \citet{lorenz_etal_20} found there are two distinct groups of inner bars in terms of their in-plane length and ellipticity. Whether the properties of long-lived inner bars are systematically different from those of recurrent ones needs further investigation.

%-----------
%-- Sect. 6
%-----------

\section{Summary}
\label{sec:summary}

In this study, we have used high-resolution numerical simulations to explore gas flow patterns in a galaxy with two independently rotating bars (S2Bs). Different gas physics setups are tested, and the gravitational effect of the inner bar is highlighted with control models. Our main findings are summarized as follows:

(1) We find the inner bar can drive gas flow inwards. The inflow is periodic, and is stronger when the two bars are perpendicular with respect to each other, possibly due to more clear shocks in the inner bar at this orientation (Figure~\ref{fig:S2Bisosmuinflow}). The time-averaged central inflow rate in our S2B model is about 3 times higher than that in the single-barred one (Figure~\ref{fig:SBssfrinflow}).

(2) Gas forms multiple streams down to the center from the nuclear ring in our S2B model, while this is not observed in the single-barred simulation (Figure~\ref{fig:S2BS1Bden}). Outside the nuclear ring, the flow patterns are very similar between single- and double-barred galaxies. A gas inner bar is formed in the S2B model without stellar feedback, but it is erased when feedback is included (Figure~\ref{fig:SBsdencomp}). The gas flow pattern in our S2B model with {\tt SMUGGLE} enabled is qualitatively consistent with a few observed S2Bs (Figure~\ref{fig:ngc1433}).

(3) The S2B model has the highest star formation rate compared with the single- and non-barred models (Figure~\ref{fig:SBssfrall}) at a given stellar mass. This may be explained by a higher star formation efficiency (or a shorter depletion time), together with a higher accumulated gas mass in the center of the S2B model. We also find that the depletion time in the central region is shorter than that in the disk, similar to recent observations (Figure~\ref{fig:SBsKSplot}).

\software{
{\tt AREPO} \citep{spring_10,weinbe_etal_20},
NumPy \citep{2020NumPy-Array},
SciPy \citep{2020SciPy-NMeth},
Matplotlib \citep{4160265},
Jupyter Notebook \citep{Kluyver2016jupyter}
}

\begin{acknowledgments}
We thank the anonymous referee for suggestions that help to improve the presentation of the paper. 
We thank Feng Yuan and Mou-Yuan Sun for insightful discussions. The research presented here is supported by the National Natural Science Foundation of China under grant No. 12103032, and is partially supported by the National Key R\&D Program of China under grant No. 2018YFA0404501; by the National Natural Science Foundation of China under grant Nos. 12025302, 11773052, 11761131016; by the ``111'' Project of the Ministry of Education of China under Grant No. B20019; and by the Chinese Space Station Telescope project. M.D. acknowledges the support of the Fundamental Research Funds for the Central Universities. J.S. acknowledges the support of a \textit{Newton Advanced Fellowship} awarded by the Royal Society and the Newton Fund. A.B. was supported by the Future Investigators in NASA Earth and Space Science and Technology (FINESST) award number 80NSSC20K1536 during the completion of this work.
This work made use of the Gravity Supercomputer at the Department of Astronomy, Shanghai Jiao Tong University, the facilities of the Center for High Performance Computing at Shanghai Astronomical Observatory, and the High-performance Computing Platform of Peking University.
\end{acknowledgments}

\bibliographystyle{aasjournal}
\bibliography{gasdynamics}

\end{document}